\documentclass[12pt]{article}
\usepackage[margin=1in]{geometry}
\usepackage{fancyhdr}
\usepackage{latexsym}
\usepackage{amsmath}
\usepackage{amsfonts}
\usepackage{amssymb}
\usepackage{psfrag}
\usepackage{graphicx}
\usepackage{setspace}
\usepackage{amsthm}
\usepackage{multirow}
\usepackage{booktabs}
\usepackage{float}
\usepackage[table,xcdraw]{xcolor}
\usepackage{times} 
\usepackage{array}
\usepackage{makecell}
\usepackage{graphicx}
\usepackage{bbm}

\providecommand{\keywords}[1]
{
  \small	
  \textbf{\textit{Keywords---}} #1
}

\begin{document}

\title{\hrulefill\\A Framework for Incorporating Behavioural Change into Individual-Level Spatial Epidemic Models\\\hrulefill}
\author{Madeline A. Ward$^1$\footnote{\textit{Corresponding author:} madeline.ward1@ucalgary.ca} \and Rob Deardon$^{1,2}$ \and Lorna E. Deeth$^3$ }
\date{\small $^1$ Department of Mathematics and Statistics, University of Calgary, Calgary, AB \\
$^2$ Faculty of Veterinary Medicine, University of Calgary, Calgary, AB \\
$^3$  Department of Mathematics and Statistics, University of Guelph, Guelph, ON \\
}

\maketitle

\begin{abstract}
During epidemics, people will often modify their behaviour patterns over time in response to changes in their perceived risk of spreading or contracting the disease. This can substantially impact the trajectory of the epidemic. However, most infectious disease models assume stable population behaviour due to the challenges of modelling these changes. We present a flexible new class of models, called behavioural change individual-level models (BC-ILMs), that incorporate both individual-level covariate information and a data-driven behavioural change effect. Focusing on spatial BC-ILMs, we consider four ``alarm" functions to model the effect of behavioural change as a function of infection prevalence over time. We show how these models can be estimated in a simulation setting. We investigate the impact of misspecifying the alarm function when fitting a BC-ILM, and find that if behavioural change is present in a population, using an incorrect alarm function will still result in an improvement in posterior predictive performance over a model that assumes stable population behaviour. We also find that using spike and slab priors on alarm function parameters is a simple and effective method to determine whether a behavioural change effect is present in a population. Finally, we show results from fitting spatial BC-ILMs to data from the 2001 U.K. foot and mouth disease epidemic. 
\end{abstract}

\section{Introduction}\label{sec:intro}

Human behaviour often has a substantial impact on the transmission dynamics of infectious diseases, whether it be by directly reducing one's own risk of contracting or passing on a disease, or by taking preventative action to protect animals or crops during an outbreak. Incorporating the effect of behaviour change (BC) into infectious disease transmission models may lead to a more accurate representation of the disease spread \cite{weitz2020}. Additionally, having the ability to identify how behaviour changes over time and how these changes impact transmission can provide valuable information to public health units and policy makers. \\

Incorporating BC into a data-driven statistical infectious disease model has the potential to improve our understanding of infectious disease transmission processes and increase the accuracy of model fit and forecasting. However, there are many challenges associated with doing this. Firstly, and particularly in historical contexts, there may be limitations in availability of the type of data which may be used to inform BC. We may expect mobility data such as Google COVID-19 Community Mobility Reports \cite{googlemobility}, livestock movement records, or mobile phone records to be informative of BC, for example. However, mobility data has only been available as of fairly recently and still has differential availability between countries \cite{wardle2023}. One data source may also not be sufficient for informing BC; with COVID-19, while mobility data may capture some changes in contact patterns, it would not capture behaviours such as physical distancing and mask-wearing which are also important factors in influencing spread \cite{mendezbrito2021}. Data for many such behaviours would likely not exist for most populations, or over an extended time period. Furthermore, even if such data is available, using it in a model would limit forecasting ability since it would only be known up to the current time point. Finally, human behaviour is a very complex process, and it is difficult to select a function by which to model it. Funk \textit{et al.} (2015) provides a discussion of some additional challenges for incorporating BC into infectious disease models.  \\

Likely due to the data-related challenges of measuring BC, studies that incorporate BC have been largely limited to deterministic models and have focused on model characteristics. Funk,  Salath{\'e} and Jansen (2010) provide an early review of mathematical models of human diseases that incorporate the effect of individual behavioural responses due to extrinsic information. In their review, they classify studies by the source of information used to inform BC (belief- or prevalence-based, considered on either a local or global scale) as well as the effect the BC has in the model (change of disease state, parameter modification, or change in contact structure). Verelst \textit{et al.} (2016) carried out a more recent systematic review of BC models, and noted that less than 20\% of included studies used any real data to fit or validate their models. Mathematical modelling studies are helpful for exploring possible mechanisms of behaviour change in theoretical scenarios, however, it is also important to investigate the practical suitability of these models in analysing and interpreting real data. \\

Most of the previous work on BC models has been done within a population-averaged framework. Population-averaged models are computationally efficient and often appropriate for large populations. However, such models assume homogeneity of susceptibility and mixing patterns across all individuals in the population. For some diseases and in smaller populations, a more complex model with less restrictive assumptions may be necessary to adequately describe transmission. Therefore in our study, we focus on the individual-level model (ILM) framework described by \cite{deardon2010}. \\

Individual-level models, alternatively known as individual-based models, can flexibly incorporate individual-level covariate information that may affect susceptibility or, when infected, transmission ability. They can also incorporate indicators of population mixing, such as the spatial locations of (or distances between) individuals or contact networks, thus relaxing the homogeneous mixing assumption made in many models. However, ILMs have generally been constructed based on the aforementioned assumption that susceptibility and population mixing behaviours remain constant over time, with temporal differences in transmission dynamics arising solely due to the progression of individuals through different infectious states. This ignores, for example, the possibility that individuals or the population overall may be modifying their infection probabilities based on their perception of risk at a given time. \\

Here, we propose a new class of ILMs, called BC-ILMs, that allows for time-varying susceptibility or transmission based on a prevalence-driven BC effect. Focusing on spatial ILMs, we investigate the properties of these BC-ILMs through a simulation study, including an exploration of the effects of misspecifying the so-called ``alarm" (or BC) function to address the challenge of selecting a suitable function to describe BC. Additionally, we suggest a method that can be used to screen a data set for the presence of a BC effect. Finally, we illustrate the use of our methods on real data from the 2001 U.K. epidemic of foot and mouth disease in livestock. 

\section{Methods}

\subsection{Behavioural change individual-level models}

Deardon \textit{et al.} (2010) introduced a framework for ILMs that allows the probability of infection over discrete time-steps to be modelled based on individual-level data, which may include individual-level covariate or spatial information. These models are typically placed within compartmental (or multi-state) frameworks, such as the susceptible, infectious, removed ($SIR$) framework. In the $SIR$ framework, individuals progress from being susceptible to the disease to becoming infectious and able to transmit the disease, and finally to being in a permanent removed state. After removal, individuals can no longer contract nor transmit the disease. At any given time-point, each individual in the population must be classified as either susceptible, infectious, or removed, and can only progress to other states in a forward direction. The transition from susceptible to infectious is determined by an infection probability, and the transition from infectious to removed is dictated by the infectious period of the disease, $\lambda_I$. The infectious period may be considered to be the same or varying across individuals, and can either be assumed to be known or be estimated as part of the model. \\

Let $S(t)$, $I(t)$, and $R(t)$ represent the sets of susceptible, infectious, and removed individuals at time $t$  respectively. The general form of an epidemic ILM is:
\begin{equation} \label{eq:ilm}
P(i, t) = 1 - \text{exp} \left[ \left\lbrace -\Omega_S(i) \sum_{j \in I(t)} \Omega_T(j)\kappa(i,j) \right\rbrace + \epsilon(i,t) \right],
\end{equation}
where $P(i,t)$ represents the probability of susceptible individual $i$ becoming infected at time $t$ (leading to them becoming infectious at time $t+1$). The $\Omega_S(i)$ term is a function of susceptibility factors describing the risk of susceptible individual $i$ contracting the disease, and $\Omega_T(j)$ is a function of transmissibility factors describing the risk of infectious individual $j$ transmitting the disease to susceptible individual $i$. The infection kernel, $\kappa(i,j)$, incorporates factors such as spatial separation between pairs of susceptible and infectious individuals. The final term, $\epsilon(i,t)$, represents infections that arise due to factors otherwise unexplained by the model. A constant $\epsilon(i,t)$, for example, might represent a constant and spatially homogeneous rate of introduction of new infections into the epidemic, originating from outside the population under consideration. \\

We propose a new class of ILM called the BC-ILM. In the BC-ILM we introduce an ``alarm function" term. The alarm function reflects the possibility that individuals modify their behaviour over time according to their perception of either becoming infected or infecting others. The alarm function $a_t$ may be included as a multiplicative factor to any of the terms of the ILM, provided that a function $f_k(\cdot)$ of $a_t$ results in $P(i,t)$ decreasing as $a_t$ increases. Therefore the BC-ILM may take one of four forms, where each form includes one of the bold terms in the following equation:
\begin{equation} \label{eq:ilm}
P(i, t) = 1 - \text{exp} \left[ \left\lbrace -\Omega_S(i; \boldsymbol{a_t})  \sum_{j \in I(t)} \Omega_T(j; \boldsymbol{a_t}) \kappa(i,j; \boldsymbol{a_t})  \right\rbrace + \epsilon(i,t; \boldsymbol{a_t})  \right].
\end{equation}
One may also consider additional forms of the BC-ILM where more than one of the bold terms are retained, if BC is expected to impact more than one aspect of disease transmission. \\

The likelihood function for a BC-ILM remains the same as for a general ILM. For an epidemic fit within an SIR framework, we can write the likelihood for the observed data consisting of the infectious states of each individual at a given time-point $t$. The likelihood is given by the product of the infection probabilities for individuals who are infected at time $t$ and so become infectious at time $t+1$, multiplied by the product of the complement of the infection probabilities for susceptible individuals at time $t$ who do not become infected, and thus remain susceptible at time $t+1$. This can be expressed as:
\begin{equation}
f_t\left(S(t), I(t), R(t) | \boldsymbol{\theta} \right) = \left[ \prod_{i \in I(t+1) \setminus I(t)} P(i,t) \right] \left[ \prod_{i \in S(t+1)} \left( 1 - P(i,t) \right) \right].
\end{equation}
The likelihood for the epidemic data $\boldsymbol{D}$ across all time-points in the study period, given values for the model parameters $\boldsymbol{\theta}$, is then given by:
\begin{equation}   \label{eq:ILMlik}
f\left(\boldsymbol{D} | \boldsymbol{\theta} \right) =  \prod^{t_{max}}_{t = 1} f_t\left(S(t), I(t), R(t) | \boldsymbol{\theta} \right).
\end{equation}

\subsection{Spatial ILMs}

We can define a spatial ILM by selecting a spatially-based infection kernel. For example, the infection kernel might be a function of the Euclidean distance between individuals. We will focus on a simple spatial ILM (SILM), referred to herein as the ``baseline SILM", where we assume each individual has equal susceptibility and transmissibility values, and the variability in infection probability between individuals at a given time point arises purely due to spatial information. Spatial location is considered to be static throughout the study period. The model for the baseline SILM is given by:
\begin{equation} \label{eq:silm}
P(i, t) = 1 - \text{exp} \left[-\alpha \sum_{j \in I(t)} d_{ij}^{-\beta} \right], \hspace{1cm} \alpha, \beta > 0.
\end{equation}  
In this model, $\Omega_S(i) = \alpha$ and $\Omega_T(j) = 1$ represents the susceptibility and transmissibility levels of all individuals, respectively. The infection kernel is $\kappa(i,j) = d_{ij}^{-\beta}$, where $d_{ij}$ is the Euclidean distance between susceptible individual $i$ and infectious individual $j$. The spatial parameter $\beta$ determines the decay of infection risk over distance. \\

The baseline SILM given in Equation \ref{eq:silm} can easily be extended to allow susceptibility or transmissibility to vary between individuals. For example, we could use a $k-$level categorical susceptibility covariate such that  $\Omega_S(i) = \alpha_{z_i}$, where $z_i \in {0, 1, ..., k-1}$ indicates to which susceptibility subgroup individual $i$ belongs. However, such models still assume that the susceptibility and transmissibility of an individual remains constant across time. We can use BC-ILMs to relax this assumption (note that while we continue to use the term BC-ILMs, for the remainder of the article this term is used to refer to spatial BC-ILMs specifically). 

\subsection{Spatial BC-ILMs} \label{sec:bc-ilms}

In this paper, we consider two options for where the BC term is added into the simple SILM. Firstly, we consider a model where BC impacts an individual's susceptibility level, with $\Omega_S(i; a_t) = \alpha(1-a_t)$. This gives a BC-ILM in the form of:
\begin{equation} \label{eq:bcilma}
P(i, t) = 1 - \text{exp} \left[ -\alpha \boldsymbol{(1-a_t)}  \sum_{j \in I(t)} (d_{ij}+1)^{-\beta} \right].
\end{equation}
Secondly, we consider the scenario where BC influences the spatial parameter, $\beta$, with $\kappa(i,j; a_t) = (d_{ij}+1)^{-\beta(1-a_t)^{-1} }$:
\begin{equation} \label{eq:bcilmb}
P(i, t) = 1 - \text{exp} \left[ -\alpha \sum_{j \in I(t)} (d_{ij}+1)^{-\beta\boldsymbol{(1-a_t)^{-1}} } \right].
\end{equation}
Note that one is added to the Euclidean distances to avoid an issue where an increased BC effect would increase transmission between susceptible and infectious individuals when $d_{ij}<1$ in Equation \ref{eq:bcilmb}. The same adjustment was made in Equation \ref{eq:bcilma} for consistency in the interpretations of $\alpha$ and $\beta$ between the two forms. Alternatively, prior to fitting the model the spatial coordinates in the data set could be rescaled such that the condition $\min(d_{ij}) \geq 1$ is satisfied. \\

As discussed in Section \ref{sec:intro}, there are many factors upon which BC may depend, however there are many challenges to finding and using external BC-related data. To avoid these challenges, we propose using alarm functions that depend on (reported) prevalence as a surrogate for the population's perceived risk over time. Several previous studies have illustrated the properties of a prevalence-driven BC effect in various ways in a deterministic modelling setting \cite{perra2011, kassa2011, kassa2018, eksin2019, kim2020}. \\

There are many forms of a prevalence-driven $a_t$ that may be considered; here we describe four possibilities. A simple option is to use a \textit{threshold alarm}, 
$$a_t = \begin{cases} \delta_1 & \text{if } |I_{t-1}| > \delta_2 \\ 0 & \text{otherwise} \end{cases} $$
where $0 < \delta_1 <1$ is the alarm parameter,  $|I_{t-1}|$ is the prevalence at time $t-1$, and $\delta_2 > 0$ is a prevalence threshold above which the population enters an alarmed state. With this alarm function, we assume that individuals maintain a baseline susceptibility or transmissibility level unless the prevalence is above a certain threshold, in which case they immediately reduce their susceptibility or transmissibility by a constant amount until prevalence returns to being under the threshold. This might be appropriate for modelling a situation where public health mandates are introduced based on case counts, for example. \\

We can also consider smooth functions of prevalence for our alarm function, where the behaviour change effect becomes stronger as prevalence increases. An example of a one-parameter smooth alarm function is the exponential cumulative distribution function, or \textit{exponential alarm}, $$a_t = 1-\exp({-\delta_1|I_{t-1}|}),$$ 
where $a_t$ will approach an asymptote of one at a rate determined by $0<\delta_1<1$. To make this alarm function more flexible, we can scale it using an additional parameter, $$a_t = \delta_2(1-\exp({-\delta_1|I_{t-1}|})), $$ 
where $0<\delta_2\leq1$ is the value of the asymptote (we call this the \textit{scaled exponential alarm}). The final alarm function option we consider here is a \textit{Hill-type alarm}, which has previously been used by \cite{kassa2011}, and is based on the Hill function \cite{hill1910} 
$$a_t = \frac{p_{t-1}^{\delta_2}}{\delta_1^{\delta_2} + p_{t-1}^{\delta_2}}.$$
Here $p_{t-1}$ is the proportion of the population that is infectious at time $t-1$, $0<\delta_1<1$ is the proportion of the population that needs to be infectious to achieve half the maximal BC effect, and $\delta_2>0$ is the Hill coefficient. As $\delta_2$ increases, the Hill function will produce a more gradual introduction of a BC effect than is possible with the other two smooth alarm functions mentioned here. At very large values of $\delta_2$, the Hill alarm is approximately equal to a threshold alarm.

\section{Simulation Study} \label{sec:sim_study}

\subsection{Simulation methodology}

\begin{figure}[H]
\centering
\includegraphics[scale=0.45]{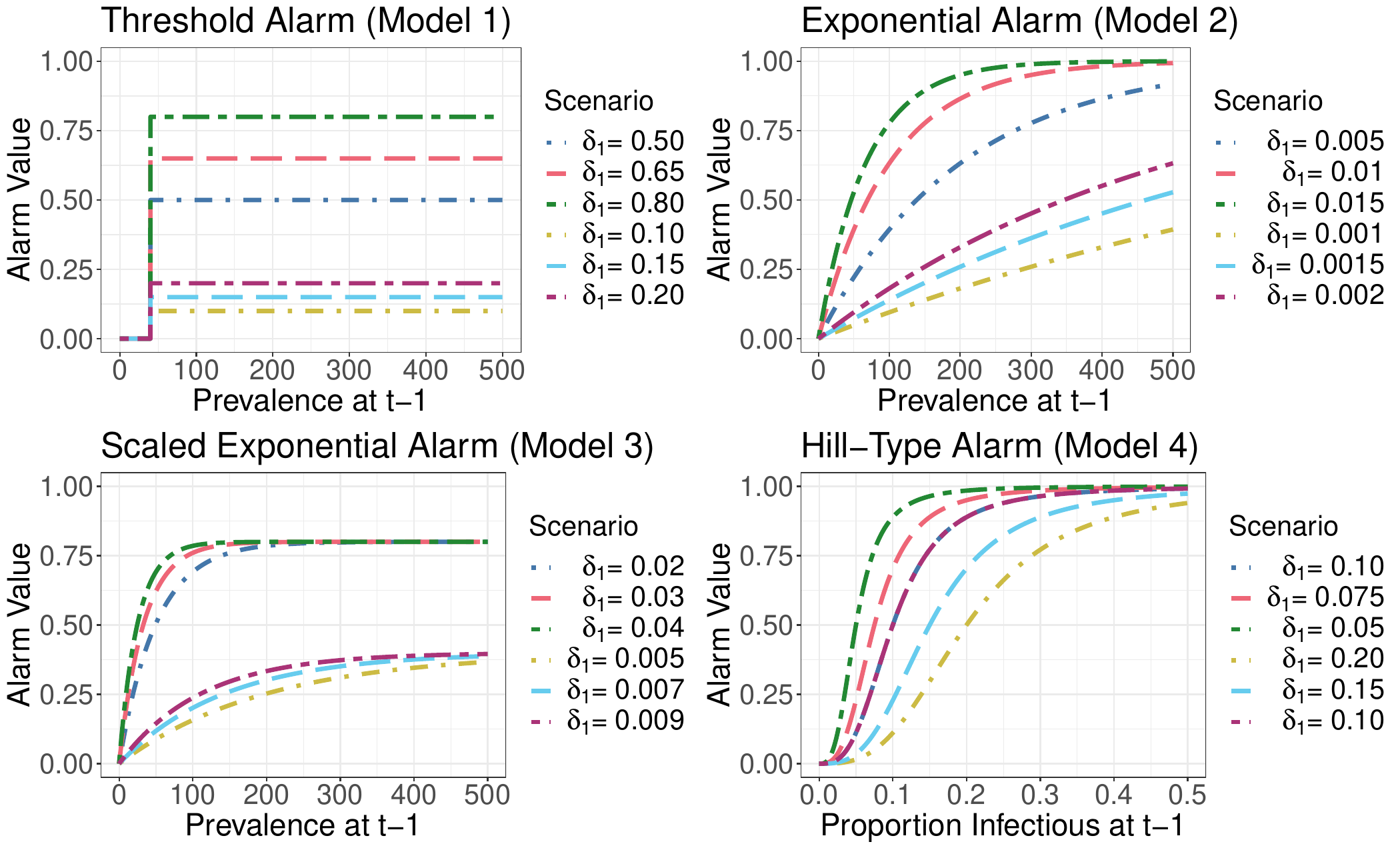}
\caption{Alarm functions for each simulation scenario. Both model types, A and B, are plotted on the same graph for a given model. Refer to Table \ref{tab:simsettings}.} \label{fig:alarms}
\end{figure}

A simulation study was conducted to explore how well BC-ILM parameters are estimated under various scenarios. We considered the four alarm functions described previously - one with a changepoint structure and three smooth functions of prevalence. These alarm functions are incorporated into one of two possible models: Model Type A which follows Equation \ref{eq:bcilma}, and Model Type B which follows Equation \ref{eq:bcilmb}. Therefore, in total we have eight BC-ILMs. Data were simulated from each model under three scenarios: ``weak", ``medium", and ``strong" BC effects (note that these terms are used relative to each other; the ``weak" effect is still fairly strong in that it changes the shape of the epidemic curve considerably compared to the no BC case). The alarm functions and corresponding simulation settings are summarized in Table \ref{tab:simsettings}. Figure \ref{fig:alarms} shows plots of each alarm function used. A total of $m = 20$ populations each containing $n=1000$ individuals each were generated for each scenario. The spatial coordinates for individuals in each population were randomly sampled over a $200\times 200$ unit square, i.e. $x, y \sim \text{Uniform}(100,300)$. Epidemics were generated within an SIR framework across a total of $t_{\text{max}}+1=31$ time points. Three individuals from each population were selected at random to initialize the epidemics by starting their infectious periods at $t=1$.  \\

Models were fit within a Bayesian framework using Markov Chain Monte Carlo (MCMC). An adaptive random-walk Metropolis-Hastings (RWMH) algorithm was used to update parameters \cite{roberts2009}. The RWMH algorithm was run for 100,000 iterations, with the first 10,000 iterations discarded as burn-in, and convergence was assessed both visually and using Geweke's diagnostic \cite{geweke1992}. For all models and scenarios, vague Uniform(0,100) priors were used for the susceptibility and spatial (transmission) parameters ($\alpha$ and $\beta$). Vague or weakly informative priors were used for the alarm function parameters (see Table \ref{tab:priors}). The MCMC computations were performed using the Julia language version 1.6.3 in the University of Calgary's Advanced Research Computing (ARC) network \cite{bezanson2017}. Post-model fitting analyses and plotting were performed in R \cite{rlang}.\\

To evaluate model fit, we considered the posterior predictive distribution (PPD) of the epidemic curve: $\mathcal{C} = \left\lbrace | I(t+1) \setminus I(t) | \right\rbrace _{t=1}^{t_{max}+1}$, which is the set of the number of new infections at each time point throughout the epidemic period. The posterior predictive distribution of $\mathcal{C}$ was estimated by simulating epidemics from a random sample of 100 values from the MCMC-based posterior estimate. From the posterior predictive distributions we calculated 95\% highest posterior density intervals (HPDIs) for each time point of $\mathcal{C}$. We also assessed each model's forecasting ability by fitting the model to only the first eight time points of each epidemic, and then used the resultant parameter estimates to generate 95\% HPDIs for the epidemic curve over times $t=9$ to $t=21$. For most scenarios and populations, $t=8$ is near the turning point of the largest infection peak, where the BC effect is strongest. Additionally, in all the scenarios that use the threshold alarm function, the prevalence in each population reached the threshold by $t=8$.

\begin{table}[H] 
\singlespacing
\centering \caption{Simulation study settings. Note $|I_{t-1}|$ indicates the number of infectious individuals and  $p_{t-1}$ denotes the proportion of infectious individuals at time $t-1$. The alarm function parameters are listed from top to bottom in the order of weak, medium, and strong BC scenarios.} \label{tab:simsettings}
\vspace{0.5cm}
\begin{tabular}{llccccc} 
\toprule
&  &  & \multicolumn{2}{c}{\makecell{Susceptibility\\ and Spatial\\ Parameters}}  & \multicolumn{2}{c}{\makecell{Alarm Function\\ Parameter(s)}}  \\ 
\cmidrule(r{4pt}){4-5} \cmidrule(l){6-7}
Model & Alarm Function   & \multicolumn{1}{c}{Model Type} & $\alpha$ & $\beta$ & $\delta_1$ & $\delta_2$ \\ \midrule
\multirow{6}{*}{\textbf{1}} & \multirow{6}{*}{\makecell{Threshold Alarm \\ \vspace{0.05cm} \\ $a_t = \begin{cases} \delta_1 & \text{if } |I_{t-1}| > \delta_2 \\ 0 & \text{otherwise} \end{cases} $}} & \multirow{3}{*}{A}  & \multirow{3}{*}{$2.2$}    & \multirow{3}{*}{$2.0$}     &  0.50  &  40   \\
  & &         &          &          &     0.65      &   40      \\
  & &         &          &          &     0.80      &    40     \\ \cmidrule(l){6-7}
  & & \multirow{3}{*}{B}   & \multirow{3}{*}{$2.2$}   & \multirow{3}{*}{$2.0$}   & 0.10 & 40 \\
 &  &         &          &          &    0.15       &    40     \\
  & &         &          &          &        0.20   &     40    \\ \midrule
 \multirow{6}{*}{\textbf{2}} & \multirow{6}{*}{\makecell{Exponential Alarm \\ \vspace{0.05cm} \\ $ a_t = 1-e^{-\delta_1|I_{t-1}|} $ }} & \multirow{3}{*}{A}  & \multirow{3}{*}{$2.4$}    & \multirow{3}{*}{$2.0$}     &  0.005         &    -     \\
  & &         &          &          &     0.01      &     -    \\
  & &         &          &          &    0.015       &     -    \\ \cmidrule(l){6-7}
&   & \multirow{3}{*}{B}   & \multirow{3}{*}{$2.4$}   & \multirow{3}{*}{$2.0$}   & 0.001 & - \\
 &  &         &          &          &      0.0015     &     -    \\
  &  &         &          &          &      0.002     &     -    \\ \midrule
\multirow{6}{*}{\textbf{3}} &   \multirow{6}{*}{\makecell{Scaled Exponential Alarm \\ \vspace{0.05cm} \\$a_t = \delta_2(1-e^{-\delta_1|I_{t-1}|}) $}}  & \multirow{3}{*}{A}  & \multirow{3}{*}{$2.4$}    & \multirow{3}{*}{$2.0$}     &     0.02      &     0.80     \\
  & &         &          &          &     0.03      &    0.80     \\
  & &         &          &          &     0.04      &     0.80    \\ \cmidrule(l){6-7}
 &  & \multirow{3}{*}{B}   & \multirow{3}{*}{$2.4$}   & \multirow{3}{*}{$2.0$}   & 0.005 & 0.40 \\
 &  &         &          &          &   0.007        &    0.40     \\
&   &         &          &          &     0.009      &    0.40     \\ \midrule
\multirow{6}{*}{\textbf{4}} &   \multirow{6}{*}{ \makecell{ Hill-Type Alarm \\ \vspace{0.05cm} \\ $a_t = \frac{p_{t-1}^{\delta_2}}{\delta_{1}^{\delta_2} + p_{t-1}^{\delta_2}}$ }} & \multirow{3}{*}{A}  & \multirow{3}{*}{$2.4$}    & \multirow{3}{*}{$2.0$}     &      0.10     &     3    \\
 &  &         &          &          &       0.075    &   3      \\
 &  &         &          &          &    0.05       &   3      \\ \cmidrule(l){6-7}
  & & \multirow{3}{*}{B}   & \multirow{3}{*}{$2.4$}   & \multirow{3}{*}{$2.0$}   & 0.20  & 3\\
 &  &         &          &          &     0.15     &   3      \\
 &  &         &          &          &     0.10     &      3   \\ \bottomrule
\end{tabular}
\end{table}

\begin{table}[H]
\centering \caption{Priors used on alarm function parameters in the simulation study.} \label{tab:priors}
\vspace{0.5cm}
\begin{tabular}[c]{@{}lm{8em}m{8em}@{}}
\hline
Model &  Prior for $\delta_1$ &  Prior for $\delta_2$ \\ \hline
1 (Threshold Alarm) & $\text{Beta}(1,1)$ & $\text{Gamma}(3,20)$ \\
2 (Exponential Alarm) &  $\text{Beta}(1,2)$ & - \\
3 (Scaled Exponential Alarm) & $\text{Beta}(1,2)$  & $\text{Beta}(1,1)$\\
4 (Hill-type Alarm) & $\text{Gamma}(2,4)$ & $\text{Beta}(1,2)$\\ \hline
\end{tabular}
\end{table} 

\begin{figure}[H]
\centering
\includegraphics[scale=0.65]{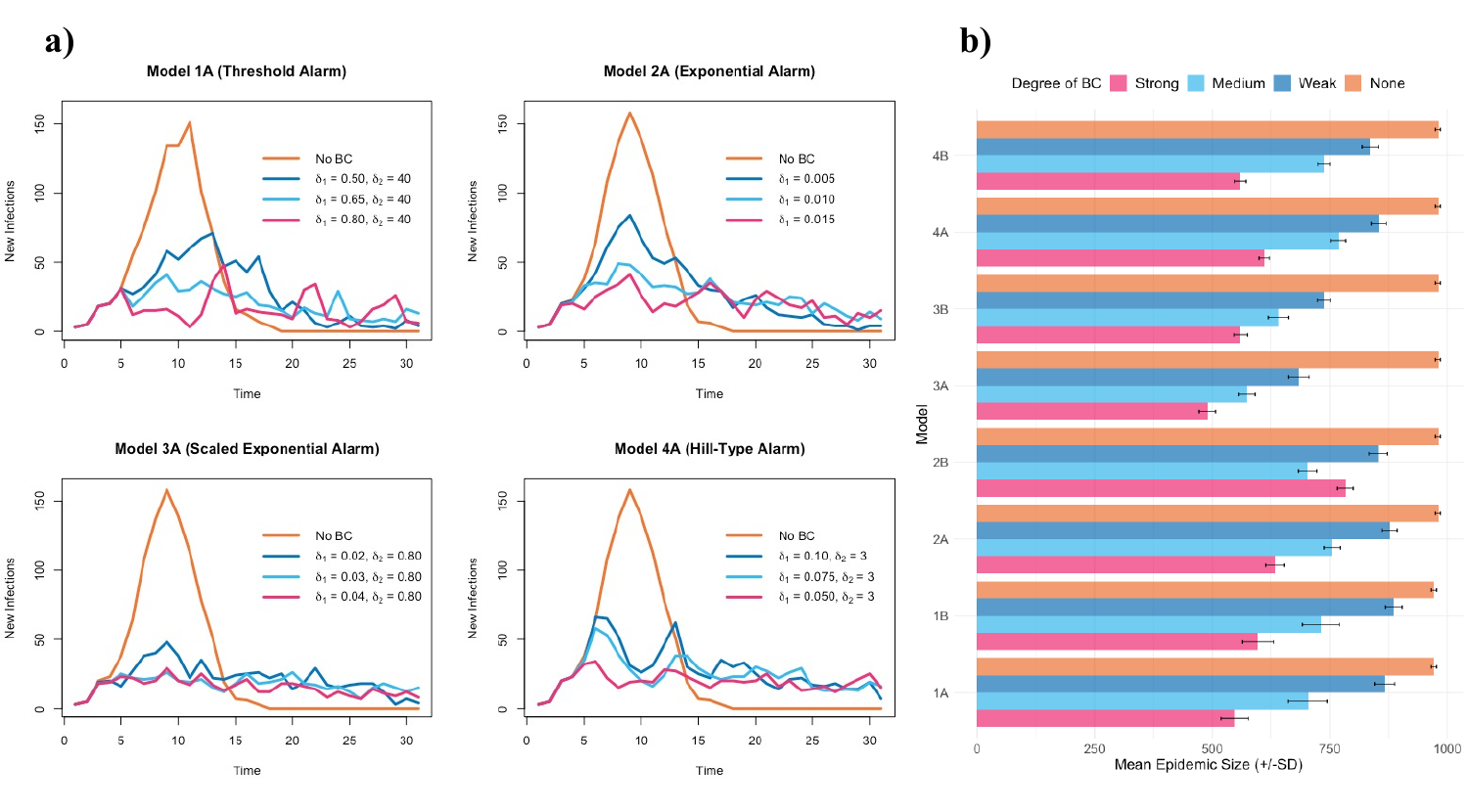}
\caption{a) Example epidemic curves under various BC scenarios for each model type. b) Average number of infections (epidemic size) over $t=1-31$ under various scenarios, with the error bars showing standard deviation in epidemic size.} \label{fig:spatial_epi}
\end{figure}

\subsection{Simulation results}

\subsubsection*{Parameter estimates}

Figures \ref{fig:paramsmod1a} - \ref{fig:paramsmod4a} show the 95\% HPDIs and posterior medians for the parameter estimates obtained under each model. Overall, parameter estimation across models was fairly good. For all models, although the posterior medians for $\alpha$ were generally lower than the true value, the 95\% HPDIs captured the true value in most populations. The posterior medians for $\beta$ consistently underestimated the true parameter value. In fact, in a number of cases, the upper limits of the 95\% HPDIs for $\beta$ were also lower than the true value. In most cases, the 95\% HPDIs for the alarm function parameters ($\delta_1$ and $\delta_2$) captured the true parameter value. When looking at posterior median estimates, it was more common to see underestimation of  $\delta_1$ in Models 1, 2 and 3, and $\delta_2$ in Model 3. These parameters, when lowered, decrease the BC effect and increase the rate of epidemic spread. Conversely, the remaining alarm function parameters which, when lowered, increase the BC effect and slow epidemic spread, were often overestimated. \\

A lower value of $\alpha$ results in a slower spreading epidemic, as it indicates that individuals are at lower susceptibility to the disease. Since a BC effect also results in an overall slower spreading epidemic, the downward bias observed in the estimates for $\alpha$ may be a result of compensation for the BC effect. This aligns with the patterns observed in the alarm function parameter estimation, as the parameters which tended to be somewhat underestimated were those that decrease the BC effect when lowered, and the parameters which tended to be overestimated were those that decrease the BC effect when raised. Therefore, in general, the BC effect was underestimated a little in most populations, which (if not compensated for by the susceptibility and spatial parameters) would lead to a faster spreading epidemic.  \\

As the value of $\beta$ decreases, susceptible individuals do not need to be as close in distance to infectious individuals to become infected. Decreasing the value of $\beta$ thus results in a more ``randomly" spreading epidemic as infectious individuals can infect susceptibles from further away. Because of this, it is common to see $\beta$ underestimated when fitting such spatial ILMs (see for example Malik, Deardon and Kwong, 2016 and Mahsin, Deardon, and Brown, 2022). In addition to being more spatially random, epidemics generated from a model with a lower value of $\beta$ tend to spread faster. The underestimation of $\alpha$ and the BC effect also therefore help compensate for the underestimation of $\beta$. Due to this correlation between all the parameters, estimation of the parameters for ILMs can be quite difficult, so overall the estimates we have observed here can be considered reasonably good for this type of model.

\subsubsection*{Posterior predictive ability}

\begin{figure}[H]
\centering
\includegraphics[scale=0.5]{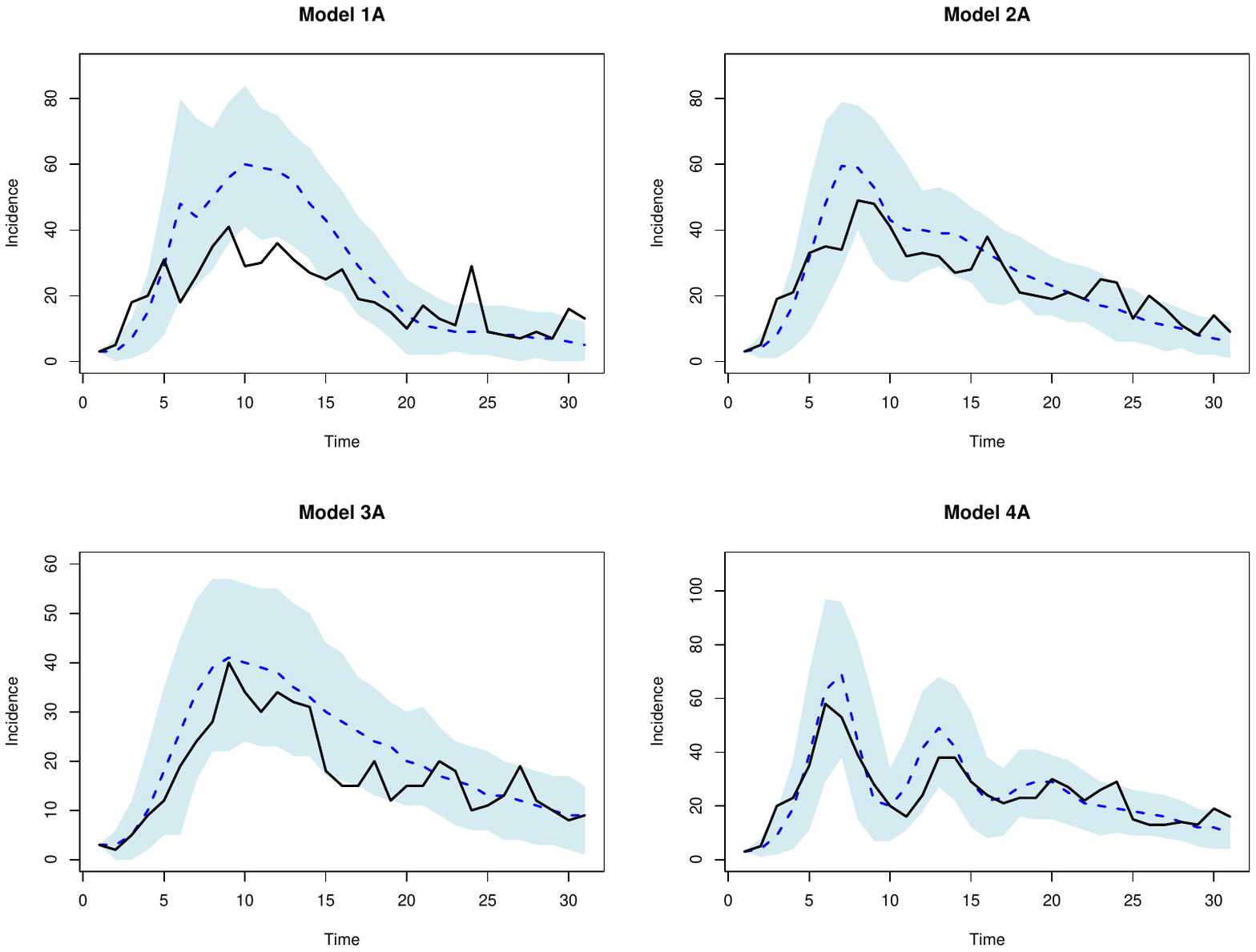} 
\centering
\caption{95\% HPDIs for the epidemic curve of one representative epidemic from the medium BC scenarios, where the true model was fit. Posterior medians are shown as dashed blue lines, and the true epidemic curves are shown with solid black lines.} \label{fig:truecurvehpds}
\end{figure}

We examined the PPD of the epidemic curve to assess the predictive performance of each model. Figure \ref{fig:truecurvehpds} shows the results for one randomly selected epidemic under each model for the medium BC scenario. Generally, the PPD of the epidemic curve indicated a good fit from each BC-ILM. The PPD of the epidemic curves under Model 1A tended to be further from the true curves when compared to the other models. In particular, in the weak and medium BC scenarios the number of incident infections at the peak of the epidemic was commonly overestimated in Model 1A. In the strong BC scenario, the epidemic data generated by Model 1A tended to have multiple peaks of similar magnitudes over the study period, and the 95\% HPDIs usually captured these peaks well. \\

For every population and scenario, the 95\% HPDIs from Model 2A captured the true curve at nearly every time point. In Model 3A, it was somewhat more common for the 95\% HPDIs to miss several time points (typically by slightly overestimating the number of incident infections near the middle of the study period). However, in the majority of populations the 95\% HPDIs captured the true curve at nearly every time point. In Model 4A, the shape of the epidemic curve tended to be captured well by the 95\% HPDIs and posterior median. In some populations, although the general shape was correct, the posterior medians were shifted relative to the true curve such that the peaks might be predicted to occur one time point too early, for example. Regardless, the 95\% HPDIs still captured the majority of time points of the true epidemic curve.

\subsubsection*{Forecasting ability}

\begin{figure}[H]
\centering
\includegraphics[scale=0.5]{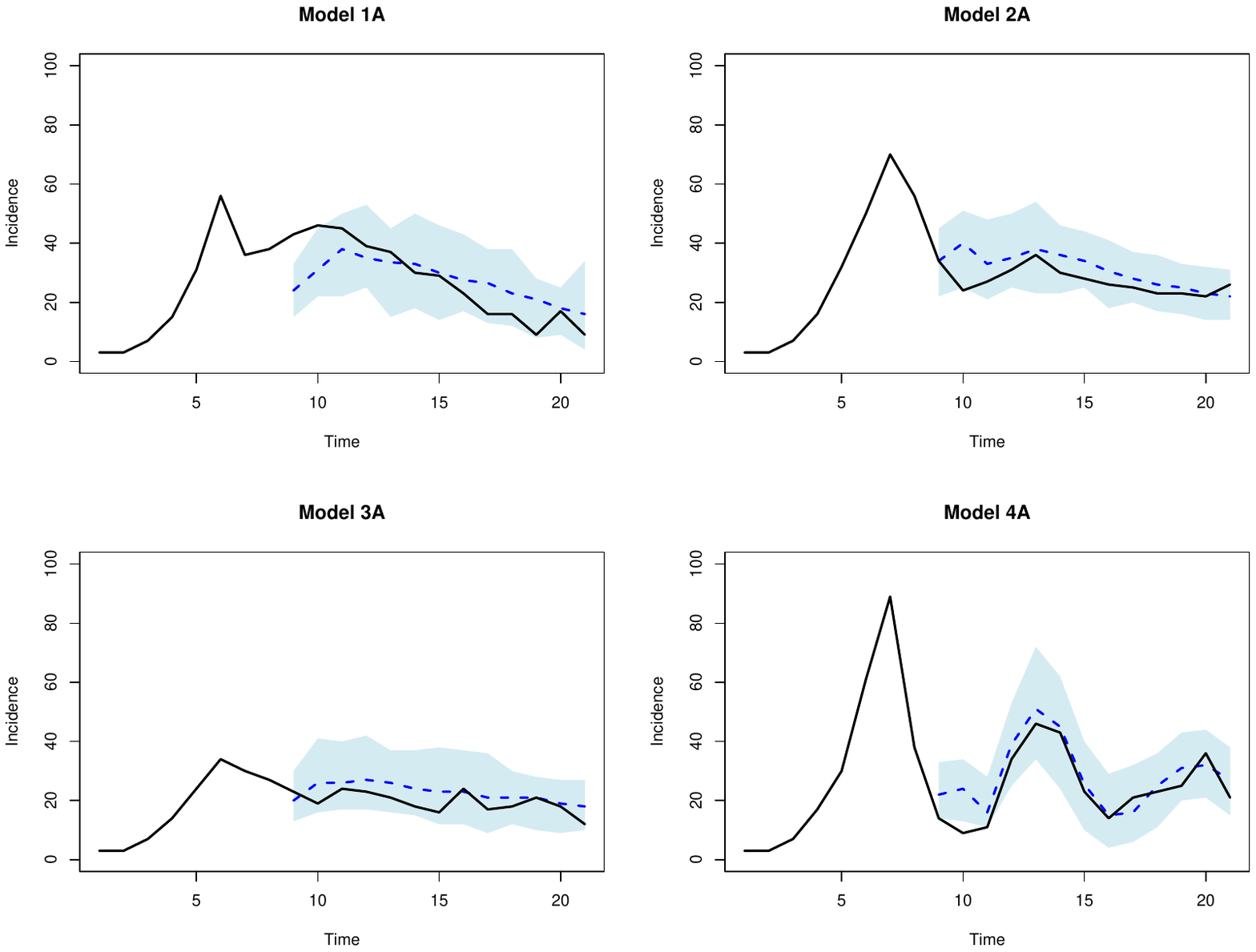} 
\centering
\caption{95\% HPDIs for the forecasted epidemic curve of one representative epidemic from the medium BC scenarios. Posterior medians are shown as dashed blue lines, and the true epidemic curves are shown with solid black lines.} \label{fig:trueforehpds}
\end{figure}

We assessed the forecasting ability of the BC-ILMs by fitting the model to the first eight time points of each epidemic and then using those posterior estimates to generate epidemics across time points $t=9-21$. Figure \ref{fig:trueforehpds} shows the posterior predictive forecasts for one randomly generated epidemic under each model for the medium BC scenario. \\

The 95\% HPDIs of the forecasts for Model 1A were generally very accurate in the weak and medium BC scenarios, with most time points of the true epidemic curves falling between the limits of the HPDIs. In the strong BC scenario, the epidemic curves generated by Model 1 had multiple peaks which were not captured as well in the forecasts. The 95\% HPDIs for the strong BC scenario forecasts tended to be very wide, with the upper limit being close to the incidence observed during peaks and the lower limit being close to the minimum incidence across most time points. \\

The 95\% HPDIs of the forecasts for Model 2A captured most time points of the true curve across all populations and scenarios. It was somewhat more common to see time points missed in the strong BC scenario where, like in Model 1A, there were often multiple large peaks in the true epidemic curve. Most of the forecasts of Model 3A were fairly accurate, however in 6/20 of the weak BC scenario populations, 5/20 of the medium BC scenario populations, and 3/20 of the strong BC scenario populations the forecasts incorrectly predicted that the incidence would continue to increase after the true epidemic peak. In these populations, the forecasted epidemic curves had very high peaks around $t=10$ to $t=15$, instead of capturing the plateau or decline of the true epidemic curves. \\

The forecasted curves from Model 4A were typically quite accurate and tended to capture the oscillations of the epidemic curves well, particularly in the weak and medium BC scenarios. In the strong BC scenario, it was more common for the 95\% HPDIs to be wide across all the timepoints rather than capturing the individual peaks and valleys (similar to what was observed in Model 1A). Despite this, in the vast majority of populations, most time points of the true epidemic curves fell within the limits of the 95\% HPDIs of the forecasts from Model 4A. \\

\section{Alarm Function Misspecification and Model Choice}

In this section, we investigate the effects of misspecifying the alarm function in a simulation setting where the generative alarm function is known. This is important since in practice, for example at the beginning of an epidemic, there may be great uncertainty as to which form of $a_t$ would be most appropriate. All models under consideration were fit to each simulated data set. In addition to each of the models listed in Table \ref{tab:simsettings}, we also fit the baseline SILM which assumes no BC (Equation \ref{eq:silm}) to each data set. Therefore, in total nine models (Models 1A-4A, 1B-4B, and the baseline SILM) were fit to each data set. We used the Watanabe-Akaike information criterion (WAIC, Watanabe, 2010) for model comparison. For each fitted model, posterior prediction and forecasting were performed using the methods described in Section \ref{sec:sim_study}. \\

\subsection{Misspecification study results}

Figure \ref{fig:curvehpds} and Figures \ref{fig:curvehpdsmod1a} - \ref{fig:curvehpdsmod4a} show 95\% epidemic curve HPDIs from under each model, for a single representative epidemic. Overall, regardless of which of the BC-ILMs was used to simulate the epidemic data, the baseline model was not able to provide a good fit. This is especially noticeable at medium and strong levels of underlying BC, where the PPD for the epidemic curve obtained with the baseline model typically peaked much later than the true epidemic does, and did not capture the initial steep increase in incidence followed by multiple infection peaks that are characteristic of BC data. \\

Amongst the four BC-ILMs, the exponential, scaled-exponential, and Hill-type alarm models (Models 2-4) tended to produce similar PPDs for the epidemic curve for each data set. Regardless of what the true model was, the 95\% HPDIs for the epidemic curve from each of these models generally provided a good fit to the true curves. When fitting the threshold alarm (Model 1), it tended to give a somewhat worse-fitting HPDI for the epidemic curves compared to the other models. Additionally, in general when the generative model was Model 1, the resultant epidemic curves seemed to be less likely to be captured well by any model. This may be attributed to the higher variation seen in the epidemic curves that arise from this model compared to the other models. \\

The results of the simulation study suggest that, overall, it is more important to include an alarm mechanism to achieve a good model fit than to specify the exact form of the alarm function correctly, as each misspecified BC-ILM still provided a substantial improvement in fit over the baseline SILM. When choosing between different alarm functions in practice, the choice between a threshold-based versus a smooth alarm function may be more important than the choice between different formulations of smooth alarm functions. Since the threshold alarm model tended to produce quite variable epidemics that were less likely to fit well even when using the true model, in an application setting it may be better to choose a smooth alarm function unless there is specific reason to believe a threshold-based BC occurred in a population. Alternatively, it may be possible to combine the threshold and smooth alarm formulations by modelling BC using the exponential alarm but only after a certain prevalence threshold is achieved, for example. This could allow for additional flexibility in situations where there is uncertainty around what BC mechanism is most appropriate.

\begin{figure}[H]
\centering
\includegraphics[scale=0.65]{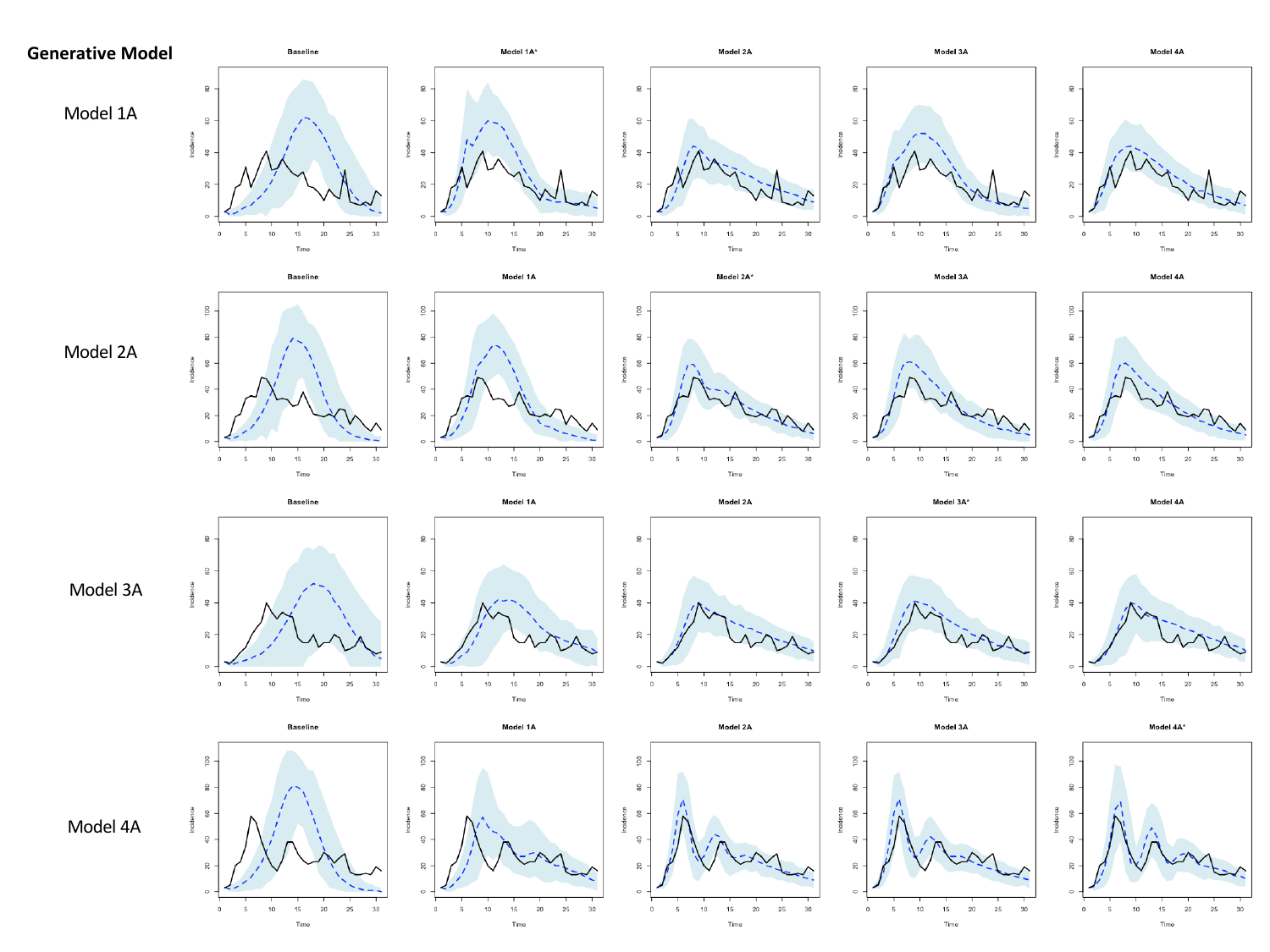} \label{fig:curvehpds}
\centering
\caption{95\% HPDIs for the epidemic curve of one representative epidemic from the medium BC scenarios. Posterior medians are shown as dashed blue lines, and the true alarm functions are shown with solid black lines. Plot titles indicate which model was fit; titles with an asterisk indicate where the true model was fit.} \label{fig:curvehpds}
\end{figure}

\begin{figure}[H]
\centering
\includegraphics[scale=0.65]{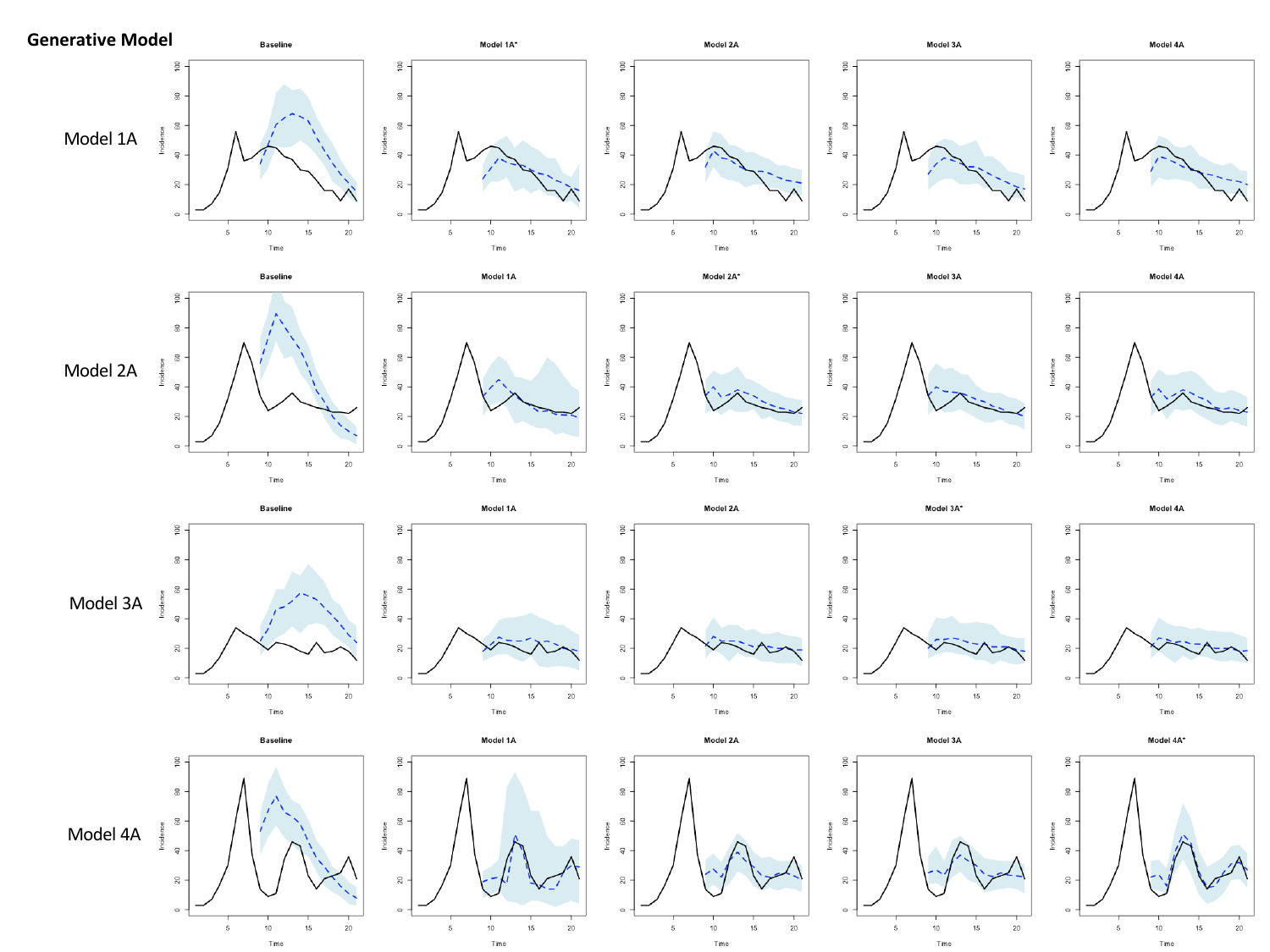}
\caption{95\% HPDIs for the forecasted epidemic curve of one representative epidemic from the medium BC scenarios. Posterior medians are shown as dashed blue lines, and the true alarm functions are shown with solid black lines. Plot titles indicate which model was fit, and row labels indicate which model was used to generate the data set. Plot titles with an asterisk show the true model fit. }
\label{fig:projcurves}
\end{figure}

Figure \ref{fig:projcurves} shows 95\% HPDIs of posterior predictive forecasts of the epidemic curve for one representative population. As was observed in the PPD of the full epidemic curves, the forecasts obtained with the baseline model were poor and considerably overestimated the incidence at the peak of the epidemics. However, the BC-ILMs all provided reasonably good forecasts for most populations. As was also seen in the PPD of the full epidemic curves, the choice of model did not have a large impact on the epidemic forecasts, although the widths of the HPDIs for Model 1 when the generative model was not Model 1 tended to be wider than those of the other three models. 

\subsection{Model choice using WAIC results}

Tables \ref{tab:propwaics} and \ref{tab:meanwaics} summarize the WAIC values obtained from fitting each model to the true data simulated under the medium BC scenario. Similar results were observed in the weak and strong BC scenarios. Table \ref{tab:propwaics} shows the proportion of time each model had the lowest WAIC value of all the models fit. When the model used to simulate the data set was Model 1, 2, or 4, the true model was most often correctly selected. However, when the generative model was Model 3, the WAIC values from Model 2 were typically lower than those from the true model. This may be because the alarm function in Model 2 is a special case of the more flexible alarm function in Model 3 where $\delta_2$ is constrained to be equal to one, such that there is one fewer parameter to estimate. The mean differences in WAIC between Models 2 and 3 (and within a given model type, eg. A) were minimal ($<2$), so in practice neither would be considered ``significantly" different from the other.\\

Generally, the difference in WAIC between types within a model (eg. comparing Model 1A to 1B) was fairly small on average (see Table \ref{tab:meanwaics}). This aligns with what was observed in the PPD for the epidemic curve, where model type typically had minimal impact on the HPDIs. The baseline model was never selected as the best model using the WAIC, and on average it had much higher WAIC values compared to when the true model was fit. Based on the WAIC and the PPD of the epidemic curve, we conclude that the baseline model is unable to provide a good fit when the data has a BC effect present.

\begin{table}[H]
\singlespacing
\centering \caption{Proportion of epidemic data sets selected using the WAIC, for the medium BC scenario. Bold numbers indicate where the true model was fit.}\label{tab:propwaics}
\vspace{0.5cm}
\begin{tabular}{cccccccccc}
 \toprule 
& \multicolumn{9}{c}{Model fit}                                                                                                                                                                            \\ \cmidrule(l){2-10}
 True Model & Base & 1A & 1B & 2A & 2B & 3A & 3B & 4A & 4B \\ 
  \midrule
1A & 0.00 & \textbf{1.00} & 0.00 & 0.00 & 0.00 & 0.00 & 0.00 & 0.00 & 0.00 \\ 
\rowcolor[HTML]{E7E6E6}
 1B & 0.00 & 0.10 & \textbf{0.90} & 0.00 & 0.00 & 0.00 & 0.00 & 0.00 & 0.00 \\ 
  2A & 0.00 & 0.10 & 0.00 & \textbf{0.60} & 0.00 & 0.05 & 0.00 & 0.20 & 0.05 \\ 
  \rowcolor[HTML]{E7E6E6}
  2B & 0.00 & 0.00 & 0.20 & 0.05 & \textbf{0.40} & 0.00 & 0.00 & 0.00 & 0.35 \\ 
  3A & 0.00 & 0.00 & 0.05 & 0.60 & 0.05 & \textbf{0.25} & 0.00 & 0.05 & 0.00 \\ 
  \rowcolor[HTML]{E7E6E6}
  3B & 0.00 & 0.05 & 0.05 & 0.10 & 0.55 & 0.00 & \textbf{0.15 }& 0.00 & 0.10 \\ 
  4A & 0.00 & 0.00 & 0.00 & 0.00 & 0.00 & 0.00 & 0.00 & \textbf{1.00} & 0.00 \\ 
  \rowcolor[HTML]{E7E6E6}
  4B & 0.00 & 0.00 & 0.00 & 0.00 & 0.00 & 0.00 & 0.00 & 0.00 & \textbf{1.00} \\ 
   \bottomrule
\end{tabular}
\end{table}

\begin{table}[H]
\singlespacing
\centering \caption{Mean difference in WAIC between the misspecified model and true model fits, for the medium BC scenario. A positive value indicates the misspecified model on average had a higher WAIC (worse fit) compared to the true model. Bold numbers indicate where the correct model number but incorrect model type (A/B) was fit.}\label{tab:meanwaics}
\vspace{0.5cm}
\begin{tabular}{cccccccccc}
 \toprule 
& \multicolumn{9}{c}{Model fit}                                                                                                                                                                            \\ \cmidrule(l){2-10}
 True Model & Base & 1A & 1B & 2A & 2B & 3A & 3B & 4A & 4B \\ 
  \midrule
1A & 35.19 & - & \textbf{5.33} & 27.05 & 29.63 & 13.16 & 18.97 & 20.34 & 36.41 \\
\rowcolor[HTML]{E7E6E6} 
  1B & 26.02 & \textbf{3.41} & - & 20.45 & 20.16 & 12.13 & 10.50 & 17.67 & 26.24 \\ 
  2A & 76.50 & 15.53 & 26.71 & - & \textbf{7.97} & 1.91 & 9.59 & 1.22 & 9.16 \\ 
  \rowcolor[HTML]{E7E6E6}
  2B & 75.06 & 24.31 & 19.10 & \textbf{4.58} & - & 6.41 & 1.33 & 4.99 & -0.31 \\ 
  3A & 40.94 & 8.87 & 11.57 & -1.05 & 4.09 & - & \textbf{5.11} & 0.62 & 5.68 \\ 
  \rowcolor[HTML]{E7E6E6}
  3B & 56.77 & 16.30 & 12.30 & 4.46 & -1.18 & \textbf{6.23} & - & 5.38 & -0.22 \\ 
  4A & 289.12 & 57.61 & 74.60 & 16.51 & 41.11 & 19.09 & 43.18 & - & \textbf{28.72 }\\ 
  \rowcolor[HTML]{E7E6E6}
  4B & 502.19 & 105.99 & 120.89 & 93.23 & 77.89 & 94.59 & 79.98 & \textbf{29.51} & - \\ 
   \bottomrule
\end{tabular}
\end{table}

\section{Spike and Slab Priors as a Behavioural Change Screening Tool}

The simulation studies in described earlier in this paper consider cases where there is a BC effect present in an epidemic data set. In scenarios where there is no BC, or only a very small degree of BC, identifiability issues can arise. For example, in the threshold alarm model, if the threshold parameter $\delta_2$ takes on a value close to or greater than the maximum prevalence observed in the data, the alarm parameter $\delta_1$ can change to any value with negligible or no effect on the value of the computed likelihood. Similarly, if $\delta_1=0$, $\delta_2$ can take on any value without affecting the likelihood. In the scaled exponential alarm model, the same phenomenon occurs when either $\delta_1 = 0$ or $\delta_2 = 0$. Additionally, although Model 2 would not have this identifiability issue since it only has one BC parameter, using a typical prior on $\delta_1$ when there is no BC present can still result in poor MCMC mixing and biased parameter estimates. \\

We propose using spike and slab priors as a screening tool for identifying whether there is a BC effect present. By taking this approach, we can identify whether a baseline SILM is adequate or if including the BC effect is necessary, then use this information to determine which type of model should be used for a given epidemic data set.

\subsection{Spike and slab priors} \label{sec:spikeslabdesc}

Spike and slab priors (also called sparsity-inducing priors) are mixture distributions with either a point mass or continuous distribution with very small variance centred at one value (the spike), and the remaining mass represented by a continuous and typically flat or non-informative distribution (the slab, $p_{\text{slab}}(\cdot)$). The mixture weights are determined by a prior probability of the variable(s) being included in the model. Spike and slab priors were originally proposed as a method for variable selection in Bayesian linear regression. In the initial discrete mixture construction of spike and slab priors by \cite{mitchell1988}, the spike follows a Dirac delta ($\delta_0$) distribution, while the slab follows a uniform distribution with a large variance. Later, \cite{george1993} proposed spike and slab priors with the construction of a continuous mixture of two normal distributions, where the spike had very small variance. In this case, effect sizes may shrink to very near but not equal to zero. Kuo and Mallick (1998) described another discrete spike and slab prior construction where indicator variables are used to determine the inclusion or exclusion of model variables \cite{kuo1998}. \\

\begin{figure}[H]
\centering
\includegraphics[scale=0.3]{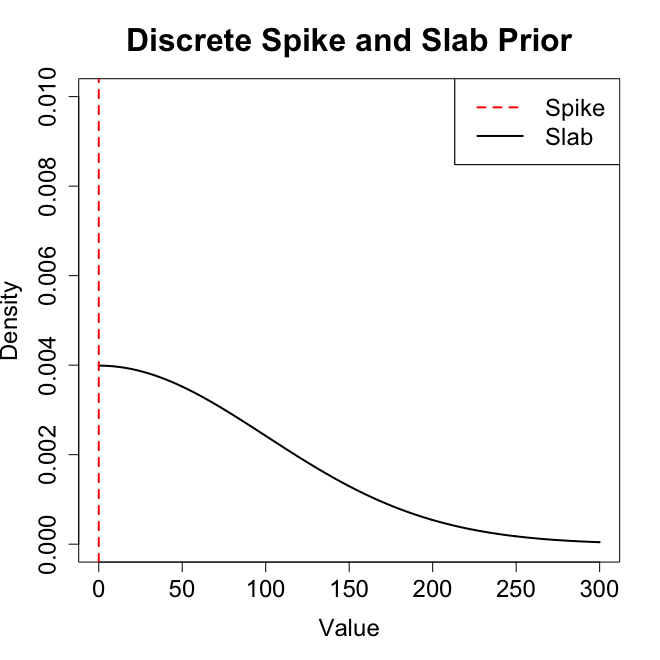}
\caption{Illustration of discrete spike and slab prior construction where $\theta_j \geq 0$. } \label{fig:spikeslab}
\end{figure}

In the case of BC-ILMs, we wish to allow the alarm function parameters to take on values of zero if there is no BC present in the system, and therefore we will focus on the discrete Kuo-Mallick construction of spike and slab priors (illustrated in Figure \ref{fig:spikeslab}). Mathematically, these priors on the $j^\text{th}$ parameter $\theta_j$ can be expressed as:
$\theta_j = z_j\theta^*_j$, where $z_j \sim \text{Bern}(\pi_j)$, $\theta^*_j \sim p_{\text{slab}}(\cdot)$, and $\pi_j$, the inclusion probability for $\theta_j$, may either be set to a constant or might follow a distribution (typically a Beta distribution). When there is little information available to inform whether $\theta_j$ should take on a non-zero value, $\pi_j = 0.5$ or $\pi_j \sim \text{Beta}(5,5)$ are often used. 

Implementation of discrete spike and slab priors for BC-ILMs presents some challenges. First, most applications of spike and slab priors have been to linear regression, where they are conjugate to the likelihood function and Gibbs sampling can be employed. This is not the case with other models such as BC-ILMs, and thus (less efficient) independence sampling from the priors themselves is generally used to update the alarm function parameters. Second, to prevent the identifiability issue described above, a joint prior for $\delta_1$ and $\delta_2$ which constrains them to both equal zero in the same iterations is necessary. Because of this requirement, independence sampling is particularly inefficient. We therefore propose using spike and slab priors as an initial ``screening" tool to select the class of ILM which can then be fit using typical priors and MH-MCMC to obtain final parameter samples and model estimates.

\subsection{Simulation set-up} \label{sec:spikeslabsim}

In the main simulation study described in Section \ref{sec:sim_study}, only fairly strong BC effects were considered as the goal was to investigate the properties of an epidemic where BC was present. In this section, the goal is to investigate whether a weak BC effect in an epidemic  is distinguishable from no BC effect, so new simulation settings were considered. For each alarm type (excluding the Hill-type alarm), epidemics were generated under six different scenarios for each alarm and model type where the six values for $\delta_1$ were equally spaced between 0 and the smallest value in Table \ref{tab:simsettings} (inclusive). Values of $\alpha = 2.4$ and $\beta = 2$ were used in each scenario, and for $\delta_2$ the values from Table \ref{tab:simsettings} were used (except in the cases where $\delta_1 = 0$, wherein $\delta_2$ was also set equal to zero). \\

Aside from the parameter values, the spike and slab simulation study followed the same set-up as the main simulation study. The priors listed in Table \ref{tab:priors} were used as the slab distribution, and we let $\pi_j \sim \text{Beta}(5,5)$. The spike and slab step using an independence sampler for $\delta_1$ and $\delta_2$ (and an MH-MCMC sampler with vague priors for $\alpha$ and $\beta$) was run for 25,000 iterations. The selected model class (either baseline SILM or BC-ILM) was then fit to the data set using MH-MCMC for 75,000 iterations. The median value of each chain from the spike and slab step was used to initialize the chains for the MH-MCMC step. Only these final 75,000 iterations (minus a 7,500 iteration burn-in period) were used for parameter estimates and posterior prediction. 

\subsection{Results}

As in the other simulation study, we will report the results from fitting model type A as similar patterns were observed for model type B. Table \ref{tab:propss} shows the proportion of populations for which the correct model was selected through the spike and slab screening process for the various simulation scenarios. In the first row, the correct model would be the baseline SILM as there was no true BC effect in the data, whereas for the other rows the BC-ILM is the correct model. For example, the value in the first row and column of the table indicates that when fitting Model 1A (threshold alarm model) to data with no BC effect, the spike and slab screening process correctly selected the baseline SILM in 95\% of populations. When there was no BC effect in the data set, the spike and slab screening process was effective at identifying that the baseline SILM is more appropriate. When the generative model had a scaled-exponential alarm function, the spike and slab screening process always correctly selected the BC-ILM, even when the BC effect was very weak. However, when the generative model had a threshold or exponential alarm, the spike and slab screening process tended to incorrectly select the baseline SILM in the weakest BC scenarios. 

      \begin{table}[H]
      \singlespacing
      \centering
\caption{Proportion of $m=20$ generated epidemics where the correct model (baseline SILM for the No BC scenario, BC-ILM for scenarios i-v) was selected through the spike and slab screening procedure in the simulation study. Scenarios are ordered from weakest BC effect (i) to strongest (v).}
      \vspace{0.5cm}
\label{tab:propss}
\begin{tabular}{lccc}
\hline
         & \multicolumn{3}{c}{\textbf{BC-ILM Alarm Type}}     \\ \cline{2-4} 
\textbf{Scenario} & Threshold & Exponential & Scaled-Exponential \\ \hline
No BC    & 0.95      & 1.0         & 0.80               \\
i        & 0.30      & 0.0         & 1.0                \\
ii       & 0.45      & 0.05        & 1.0                \\
iii      & 0.65      & 0.35        & 1.0                \\
iv       & 0.95      & 0.40        & 1.0                \\
v        & 1.0       & 1.0         & 1.0                \\ \hline
\end{tabular}
\end{table}

\begin{figure}[H]
\centering
\includegraphics[scale=0.5]{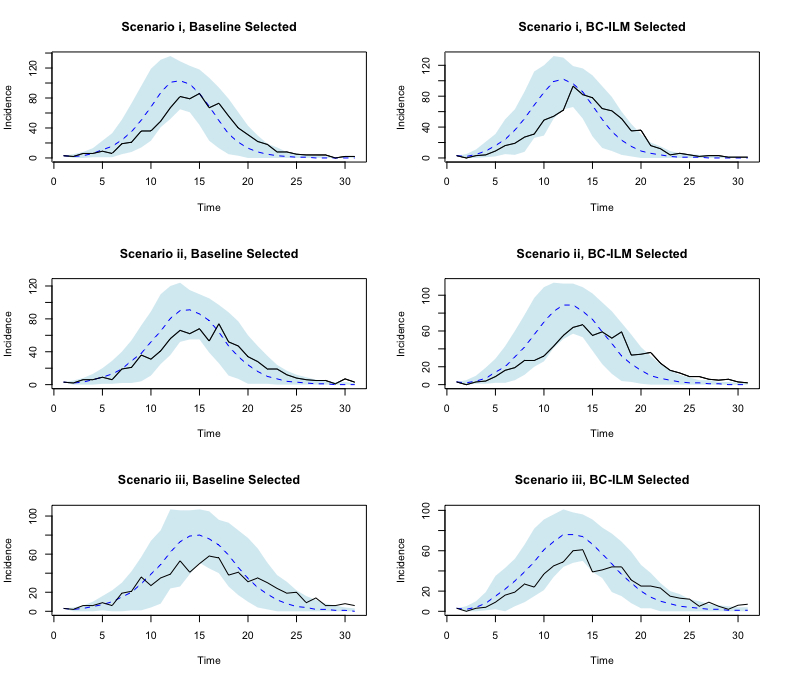}
\caption{95\% HPDIs for the epidemic curve for the three weakest BC scenarios considered in the spike and slab simulation study, where the generative model was Model 1A. The true curve is shown as a solid black line, and posterior medians are shown as a dashed blue line. The left column shows an example population where the baseline SILM was (incorrectly) selected, and the right column shows an example population where the BC-ILM was (correctly) selected. } \label{fig:sspostpred}
\end{figure}

When we examine the posterior predictive distribution of the epidemic curve for the populations where the baseline SILM was incorrectly selected, there is little difference between the abilities of the baseline SILM and the BC-ILMs to capture the true epidemic curve. Figure \ref{fig:sspostpred} shows two populations for each of the three weakest scenarios where the generative model was Model 1A: one where the true model was selected and one where the baseline SILM was selected. The 95\% HPDIs obtained by both model types are very similar for all three scenarios. By contrast, as seen in the original simulation study, when there is a stronger BC effect it becomes impossible for the baseline SILM to capture the shape of the true epidemic curve. These results show that there is a BC effect ``threshold", where for effects weaker than the threshold the baseline SILM can provide an adequate fit and the more complex BC-ILM may not be required, while for effects stronger than the threshold the BC-ILM is required to achieve a good fit.

\section{Application: 2001 U.K. Foot and Mouth Disease}

In this section we apply the methods of previous sections to a real data set from the 2001 U.K. foot and mouth disease (FMD) epidemic. During the FMD epidemic, several interventions were introduced. An overview of control strategies used is provided by \cite{kitching2005} and includes a total ban on susceptible animal movement, restrictions on veterinarian movements, and mass culling. The direct effects of these have been incorporated into previous mathematical and statistical models (see Pomeroy \textit{et al.}, 2017 for a review). However, it is possible that the general awareness of the epidemic amongst farmers, and the resultant BC, reduced transmission further in addition to the direct impacts of any control strategies. It has also been speculated that farmers decreased their adherence to control measures as cases began to decline, which may have contributed to the prolonged period of low incidence observed at the end of the epidemic \cite{ferguson2007}. We aim to use BC-ILMs to investigate whether this BC effect can be estimated from FMD data from the Cumbria region. 

\subsection{Analysis}

We used the data set described in \cite{deeth2013}, which contains 1,177 cattle and sheep farms from a portion of the Cumbria region in north-west England. The first recorded infection from the region represented in our data set occurred on day 21 of the overall epidemic. However, although we did attempt to fit the models to data starting from day 20 in the epidemic, MCMC chain convergence could not be achieved using this time period (possibly due to the sparsity of infections early on). We therefore fit our models to the infection data for days 30-50 (March 8th - 28th, 2001) of the epidemic. \\

The FMD data set contains Cartesian coordinates for each farm as well as their infection and cull dates and the number of sheep and cattle kept on the farm. A binary susceptibility covariate $z_i$ was defined based on farm size, following the methods described by \cite{deeth2013}. Farms with more than 50 cattle and/ or more than 200 sheep were considered to be at higher susceptibility to infection. Based on this definition, in our analysis 442 farms were considered to have low susceptibility ($z_i = 0$), and 735 were considered to have high susceptibility.  \\

We assume transmission takes place within an $SEIR$ compartmental framework. In an $SEIR$ framework, after a previously susceptible individual becomes infected they remain in an exposed (or latent) period before they become infectious to others. In our analysis, farms' exposed periods were considered known and fixed at $\gamma_E = 5$ days and their infectious periods were fixed at $\gamma_I = 4$ days, unless the farm was known to have been culled sooner. In the infection probability $P(i,t)$, baseline susceptibility may take on one of two values, $\alpha_0$ or $\alpha_0 + \alpha_1$, depending on the farm's covariate status. The baseline SILM infection probability is therefore given by:

$$P(i, t) = 1 - \text{exp} \left[-(\alpha_0 + \alpha_1z_i) \sum_{j \in I(t)} (d_{ij}+1)^{-\beta} \right], \hspace{1cm} \alpha_0, \alpha_1, \beta > 0.$$

For the SEIR framework, the likelihood function is changed slightly to reflect farms entering the exposed compartment instead of the infectious compartment upon infection. The likelihood function for the set of infectious states at time $t$ is given by:
\begin{equation*}
f_t\left(S(t), E(t), I(t), R(t) | \boldsymbol{\theta} \right) = \left[ \prod_{i \in E(t+1) \setminus E(t)} P(i,t) \right] \left[ \prod_{i \in S(t+1)} \left( 1 - P(i,t) \right) \right],
\end{equation*}
and the likelihood across the full study period is therefore:
\begin{equation*}   \label{eq:ILMlik}
f\left(\boldsymbol{D} | \boldsymbol{\theta} \right) =  \prod^{t_{max}}_{t_{min}} f_t\left(S(t), E(t), I(t), R(t) | \boldsymbol{\theta} \right).
\end{equation*}

We compare the baseline SILM to BC-ILMs that account for farmers changing their behaviour over time. As in the simulations study, we considered the cases where BC affects either the farms' susceptibility, or the degree to which spatial information impacts epidemic spread (ie. how nearby a susceptible farm needs to be to an infectious farm in order to have a high probability of becoming infected). In model type A, the alarm level of the population impacts susceptibility level. We multiply the entire susceptibility term by the alarm term such that the susceptibility of both low and high risk farms are scaled to the same degree by the alarm function:

$$P(i, t) = 1 - \text{exp} \left[-(\alpha_0 + \alpha_1z_i)(1-a_t) \sum_{j \in I(t)} (d_{ij}+1)^{-\beta} \right].$$ 

In model type B, the alarm level of the population impacts the spatial parameter in the same way as was described in Section \ref{sec:bc-ilms}:

$$P(i, t) = 1 - \text{exp} \left[-(\alpha_0 + \alpha_1z_i) \sum_{j \in I(t)} (d_{ij}+1)^{-\beta(1-a_t)^{-1}} \right].$$

\begin{figure}[H]
\centering
\includegraphics[scale=0.3]{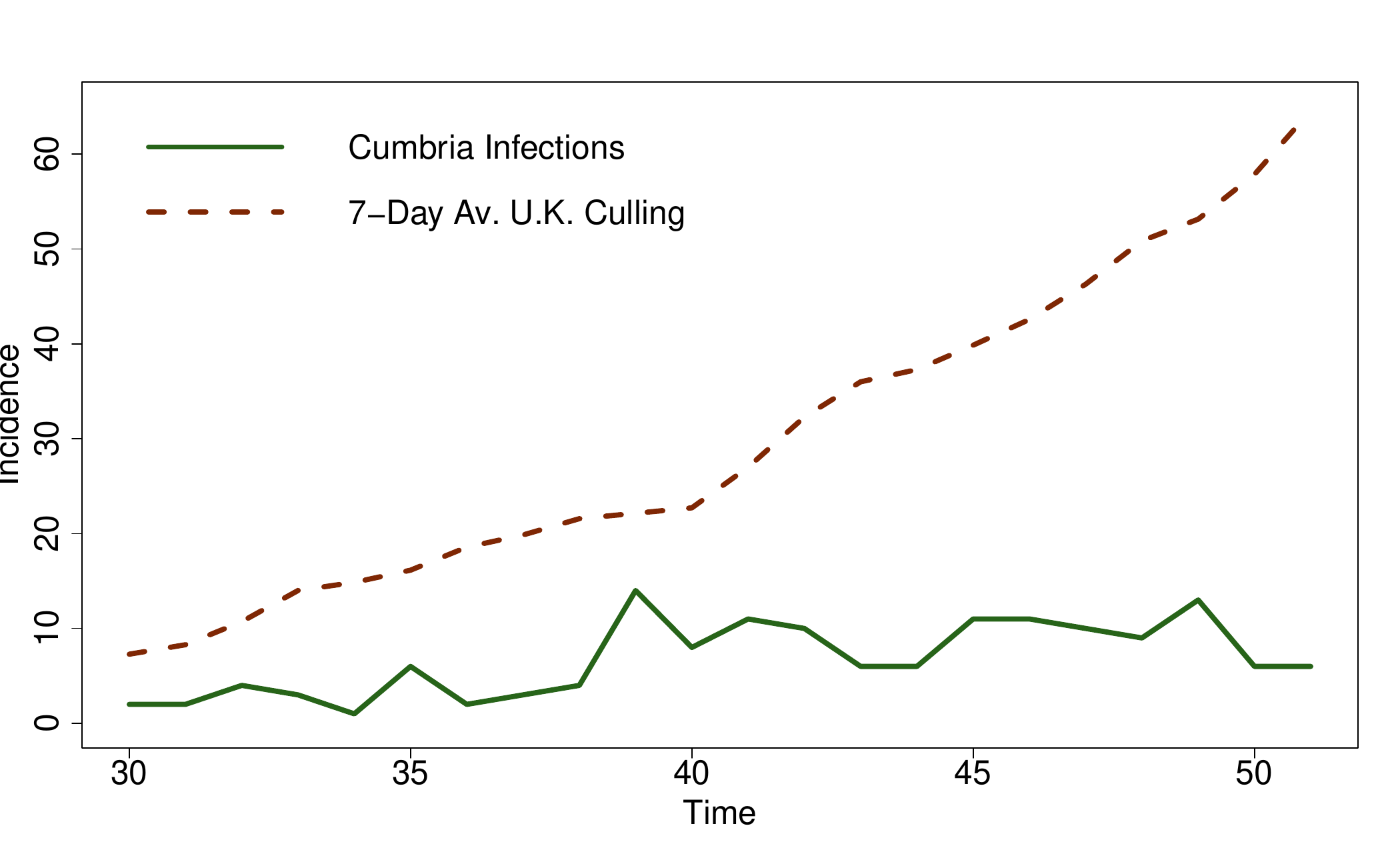}
\caption{FMD infection incidence for a subset of the Cumbria region in the U.K. from March 8th ($t=30$) - 29th ($t=51$), 2001, along with the seven day average for culling incidence in the overall U.K..} \label{fig:fmd_data}
\end{figure}

While in the simulation study we used lagged prevalence to inform the alarm function in the BC-ILMs, in the FMD analysis we found that instead using a seven day rolling average (concluding on day $t-1$) of culling incidence in the entire U.K. provided the best model fit (see Figure \ref{fig:fmd_data}). Initially, we fit the models using standard MH-MCMC with 50,000 iterations. Additionally, we fit the models again using the spike and slab screening process with 50,000 iterations in the spike and slab step and 50,000 iterations to fit the final model. In both cases, the first 5,000 iterations removed as burn-in. Vague priors were used for the susceptibility and spatial parameters ($\alpha_0$, $\alpha_1$, and $\beta$).

\subsection{Results}

MCMC convergence was not able to be achieved for Models 3 and 4 (the scaled-exponential and Hill-type alarms), so we only report the results from fitting Models 1 and 2 (the threshold and exponential alarms). Table \ref{tab:fmdparams} shows the posterior median estimates and corresponding 95\% HPDIs for each model parameter. In the BC-ILMs where alarm was assumed to affect susceptibility (Models 1A and 2A), the posterior median estimates for $\alpha_0$ and $\alpha_1$ were somewhat higher than those from the baseline SILM, though their 95\% HPDIs still captured the baseline SILM medians. The posterior median estimates for $\beta$ were very similar between the baseline SILM and Models 1A and 2A. This is expected, as in these BC-ILMs the alarm term multiplies the $(\alpha_0 + \alpha_1z_i)$ term, and thus the effect of the higher susceptibility (which leads to faster spread) in these models is counteracted by the alarm term to slow down epidemic spread during times of higher culling incidence. Similarly, in Models 2A and 2B, the posterior median estimates for $\alpha_0$ and $\alpha_1$ were very similar to those obtained using the baseline SILM, while their estimates for $\beta$ were somewhat lower than that of the baseline SILM. In this case, the faster spread resulting from a lower value of $\beta$ in the BC-ILMs would be counteracted by the alarm term multiplying it. \\

The WAIC values (Table \ref{tab:fmdwaic}) were fairly similar for the baseline SILM and Models 1A, 1B, and 2B, although the WAIC values for these BC-ILMs were all slightly lower than that of the baseline SILM. The WAIC value for Model 2A was highest, suggesting it may be a somewhat worse fit for this data compared to even the baseline SILM. Based on the WAIC, there does not seem to be much evidence that including a BC effect improves fit for this subset of the FMD data set. When we used the spike and slab screening procedure on the FMD data set, Models 1A and 2A were selected over the baseline SILM, however the baseline SILM was selected over Models 1B and 2B. There is some disagreement between the results of the spike and slab screening procedure and the WAIC values, according to which Models 1B and 2B would be selected over the baseline SILM and Model 2A would not be.  \\

\begin{figure}[H]
\centering
\includegraphics[scale=0.38]{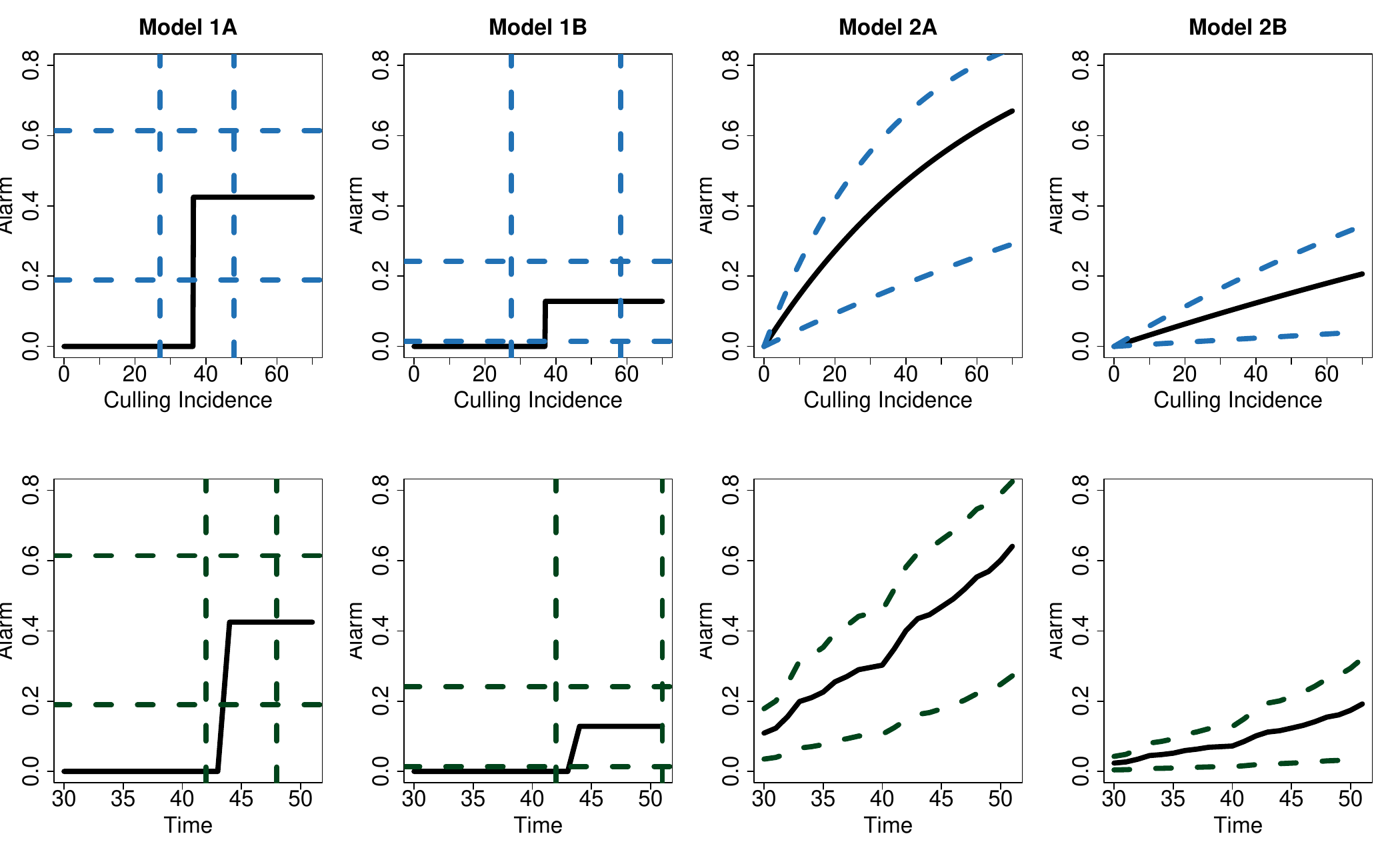}
\caption{Alarm functions estimated from each BC-ILM for the Cumbria region of the U.K.. Solid black lines represent the posterior median, while dashed lines indicate the 95\% HPDIs. Alarm value is plotted against (7-day smoothed) culling incidence in the top row and time from March 8th ($t=30$) - 29th ($t=51$), 2001 in the bottom row.} \label{fig:fmd_alarms}
\end{figure}

The estimated alarm functions (along with their 95\% HPDIs) are shown in Figure \ref{fig:fmd_alarms}. The estimated threshold alarm function suggests that alarm begins fairly late in the epidemic, with the 95\% HPDIs for when the smoothed incident culling threshold would be achieved covering $t=42$ to $t=48$, which corresponded to BC thresholds of 28.65 and 54.12 (smoothed) incident cullings, respectively. When fitting Model 1B the estimated culling incidence threshold was very similar to that obtained by fitting Model 1A, although the lower bound for the 95\% HPDI on the alarm parameter $\delta_1$ was very close to zero. The 95\% HPDI for the estimated exponential alarm function was fairly wide. Based on the wide HPDI and the higher WAIC value, it seems the exponential alarm function may not be particularly effective at describing the BC dynamics for this time period of the FMD epidemic in the Cumbria region.

\begin{table}[H]
\singlespacing
\centering
\caption{Parameter posterior median estimates and 95\% HPDIs for the baseline SILM and BC-ILMs fit to the FMD data set.} \label{tab:fmdparams}
\begin{tabular}{l
>{\columncolor[HTML]{EFEFEF}}c 
>{\columncolor[HTML]{EFEFEF}}c cc
>{\columncolor[HTML]{EFEFEF}}c 
>{\columncolor[HTML]{EFEFEF}}l }
\toprule
                & \multicolumn{2}{c}{\cellcolor[HTML]{EFEFEF}\textbf{Baseline SILM}} & \multicolumn{2}{c}{\textbf{Model 1A}}                                       & \multicolumn{2}{c}{\cellcolor[HTML]{EFEFEF}\textbf{Model 1B}}                                                 \\ \cmidrule{2-7} 
 \textbf{Parameter}   & Estimate                     & 95\% HPDI                             & Estimate                         & 95\% HPDI                                  & Estimate                                     & \multicolumn{1}{c}{\cellcolor[HTML]{EFEFEF}95\% HPDI}            \\ \midrule
$\alpha_0$           & 0.0127                       & (0.0061, 0.0249)                    & 0.0169                           & (0.0079, 0.0352)                         & 0.0136                                       & \multicolumn{1}{c}{\cellcolor[HTML]{EFEFEF}(0.0067, 0.0256)}   \\
$\alpha_1$            & 0.0592                       & (0.0322, 0.1023)                    & 0.0790                           & (0.0424, 0.1438)                         & 0.0625                                       & \multicolumn{1}{c}{\cellcolor[HTML]{EFEFEF}(0.0341, 0.1102)}   \\
$\beta$               & 2.0491                       & (1.7531, 2.3325)                    & 2.0541                           & (1.7656, 2.3369)                         & 1.9506                                       & \multicolumn{1}{c}{\cellcolor[HTML]{EFEFEF}(1.6483, 2.2442)}   \\
$\delta_1$           &                              &                                     & 0.4257                           & (0.1651, 0.6028)                         & 0.1283                                       & \multicolumn{1}{c}{\cellcolor[HTML]{EFEFEF}(0.0416, 0.3521)}   \\
$\delta_2$            &                              &                                     & 36.3380                          & (28.6478, 54.1146)                       & 36.9210                                      & \multicolumn{1}{c}{\cellcolor[HTML]{EFEFEF}(28.2905, 59.0117)} \\ \midrule
               & \multicolumn{2}{c}{\cellcolor[HTML]{EFEFEF}\textbf{Model 2A}}      & \multicolumn{2}{c}{\cellcolor[HTML]{FFFFFF}\textbf{Model 2B}}               & \cellcolor[HTML]{FFFFFF} &  \cellcolor[HTML]{FFFFFF}   \\ \cmidrule{2-5}
 \textbf{Parameter}    & Estimate     & 95\% HPDI      & \cellcolor[HTML]{FFFFFF}Estimate & \cellcolor[HTML]{FFFFFF}95\% HPDI          & \multicolumn{1}{l}{\cellcolor[HTML]{FFFFFF}} & \cellcolor[HTML]{FFFFFF} \\    \cmidrule{1-5}                                    
$\alpha_0$   & 0.0223     & (0.0047, 0.0535)     & \cellcolor[HTML]{FFFFFF}0.0133   & \cellcolor[HTML]{FFFFFF}(0.0062, 0.0255) & \multicolumn{1}{l}{\cellcolor[HTML]{FFFFFF}} &   \cellcolor[HTML]{FFFFFF}    \\
$\alpha_1$  & 0.1040   & (0.0222, 0.2021)   & \cellcolor[HTML]{FFFFFF}0.0619   & \cellcolor[HTML]{FFFFFF}(0.0347, 0.1038) & \multicolumn{1}{l}{\cellcolor[HTML]{FFFFFF}} &  \cellcolor[HTML]{FFFFFF}      \\
$\beta$       & 2.0562     & (1.2505, 2.3441)     & \cellcolor[HTML]{FFFFFF}1.8408   & \cellcolor[HTML]{FFFFFF}(1.5306, 2.1531) & \multicolumn{1}{l}{\cellcolor[HTML]{FFFFFF}} &  \cellcolor[HTML]{FFFFFF}     \\
$\delta_1$  & 0.0160    & (0.0043, 0.0284)   & \cellcolor[HTML]{FFFFFF}0.0033   & \cellcolor[HTML]{FFFFFF}(0.0007, 0.0062) & \multicolumn{1}{l}{\cellcolor[HTML]{FFFFFF}} &  \cellcolor[HTML]{FFFFFF}      \\ \cline{1-5}
\end{tabular}
\end{table}

\begin{table}[H]
\singlespacing
\centering
\caption{WAIC values for the models fit to the FMD data set.} \label{tab:fmdwaic}
\begin{tabular}{ll}
\toprule
Model         & WAIC    \\ \midrule
Baseline SILM & 1577.32 \\
Model 1A            & 1571.15 \\
Model 1B				& 1572.79 \\
Model 2A            & 1590.32 \\
Model 2B 			& 1573.95 \\
\bottomrule
\end{tabular}
\end{table}

\section{Discussion}

Understanding how a population changes their behaviour in response to an ongoing epidemic is important in order to be able to accurately project future case trajectories and appropriately plan public health responses \cite{west2020}. We aimed to explore the properties and estimability of spatial ILMs that include the effect of prevalence-driven BC. In our simulation study, we demonstrated that ILMs with a BC effect can be successfully estimated from epidemic data, and these models have good posterior predictive and forecasting ability. Additionally, we found that BC-ILMs are still capable of capturing epidemic curves fairly well when the model is misspecified, while baseline SILMs that ignore BC are unable to provide a good model fit. We identified an area of challenge for fitting BC-ILMs that could arise when using data where the population did not actually have change their behaviour in response to increasing prevalence, and proposed using spike and slab priors as a method to screen for this. Finally, we demonstrated the use of BC-ILMs on real data from the 2001 U.K. FMD epidemic. \\

The BC-ILMs that we proposed in this study were able to produce epidemics where a relatively low incidence is sustained over a long period of time, as well as epidemics with multiple peaks. These dynamics are unable to be represented by a traditional spatial ILM, where we see a single peak and faster depletion of susceptibles compared to BC-ILMs. This was demonstrated through the simulation study, where the baseline SILM tended to miss the location of, and overestimate the number of new cases, at the epidemic peak when it was fit to data generated from a BC-ILM. This result underlines the importance of having models which can account for BC, and we have found that it is possible to estimate a BC effect from data and use a BC-ILM to accurately forecast future cases, even when there are multiple epidemic peaks. \\

We used fairly simple models in our study, however there are many ways we could extend the BC-ILMs described here to include more complex dynamics. In the simulation study we considered a model with no covariates, such that each individual was assumed to experience BC in the same way. However, it is plausible that there may be either individual or group-level factors which influence the degree to which BC will impact infection transmission or susceptibility. For example, \cite{hills2021} found various factors had a significant effect on the odds of an individual adhering to COVID-19 social distancing rules in North London, U.K., in May 2020. These factors included whether an individual identified as highly vulnerable to COVID-19, and if they felt they had control over the social distancing of others. Risk perception has also been found to vary by age \cite{rosi2021}, so allowing for different age groups to have different alarm function parameters may improve model accuracy. \\

In the alarm function misspecification study, we found that there was not much difference in posterior predictive ability between the different BC-ILMs with smooth alarm functions, regardless of which alarm type was used to generate the data. We also did not observe noticeable differences in fit if the incorrect model type (A, where alarm affects susceptibility, versus B, where alarm affects spatial weighting) was fit. Overall, including a function of some kind to capture the BC effect appears to be more important for achieving a good model fit than correctly specifying the form of the BC-ILM. An area for future investigation is the potential effect of the baseline model on estimation of the BC effect, particularly when fitting more complex models. For example, if the model is specified to have BC multiply the susceptibility term, would failure to include an important susceptibility covariate result in poor estimation for the BC effect? Additionally, non-parametric alarm function estimation methods could be used as an alternative approach to defining parametric alarm functions. These more flexible estimation methods would have the advantage of not being constrained to follow a specific alarm function, and thus avoid the difficulty one might encounter in selecting an appropriate alarm function.  \\

We found that the spike and slab prior BC screening approach generally worked well for identifying when there was no BC, although it is a fairly computationally expensive process and in some scenarios tended to miss when weak BC was present. The screening process may be improved (at least in terms of computational efficiency) by using more complex MCMC methods, such as an adaptive algorithm (see Ji and Schmidler, 2013) or by updating the slab distribution based on the empirical distribution of early posterior samples. The spike and slab prior approach also has the limitation of not being appropriate for the Hill-type alarm function models and any other alarm functions where there are no parameter values at which the alarm function will stay constant at zero across all prevalences. Due to these limitations, screening methods for BC require more development, however this was not a main focus of this study. We believe that the approach we have described here offers a straightforward and fairly reliable way to identify whether fitting a BC-ILM or a baseline SILM would be more appropriate for a given data set. \\

In the application, our goal was to identify whether BC-ILMs could estimate a BC effect and describe population alarm. We found that there was some evidence of a BC effect in the FMD data, although including the BC effect did not appear to affect model fit considerably. As farmers would have been aware of culling incidence in the entire U.K., as opposed to only within the subset of the Cumbria region that we focused on for our analysis, we used the overall U.K incidence to inform the alarm function. This had the disadvantage that we were no longer using a direct epidemic metric that would be predicted by the model, and therefore there was no straightforward way to look at the posterior predictive distribution for the epidemic curve or forecast cases in Cumbria. With additional computational resources, future work could use BC-ILMs to model the epidemic across the entire U.K. so directly-predicted incidence or prevalence could be used to inform the alarm function like we described in the simulation studies. However, all posterior predictive and forecasting analyses of the FMD data are somewhat restricted by the culling that occurred during this epidemic and would need to be either ignored or modelled, and unfortunately modelling the U.K. 2001 culling policy is very difficult. \\

When using the spike and slab screening process on the FMD data set, models where BC impacted susceptibility were selected over the baseline SILM, while models where BC impacted the weighting of the spatial effect were not. If BC impacted the weighting of the spatial effect, that would imply that as farmers became more ``alarmed", farms further away in distance would have a lower probability of infecting a susceptible farm. This type of change in spatial weighting may have occurred despite it not necessarily being reflected in our models; it is possible that BC affecting spatial weighting occurred prior to March 8th and was sustained over our study period, thus being captured as part of the ``baseline" transmission dynamics over the time period that was modelled. Conversely, BC affecting susceptibility may have started later in the epidemic, or stronger measures to reduce a farm's susceptibility may have been taken as time went on and incidence increased.\\

Despite the limitations discussed in the introduction of using external data sources to inform behavioural change, integrating different levels of information into BC models, and infectious disease models more generally, is an area which needs further development \cite{kretzschmar2022, marion2022}. The prevalence-driven alarm described in this paper has the advantage of allowing for easy posterior prediction and forecasting, but is somewhat simplistic in that it ignores the many other factors that could influence an individual's behaviour. Where other data sources about behaviour change such as mobility data are available, it would be interesting to explore how these could be integrated into a BC model. For example, one might consider a model with prevalence-driven changes to susceptibility and mobility-informed changes to contact network structure (contact networks were not in the scope of this paper but have been previously incorporated into ILMs, see Almutiry and Deardon, 2020). \\

A final area for future work would be to allow for a time-varying alarm function. This would allow for time-dependent changes in protective behaviour adherence to be captured; for example, the strength of people's behavioural change may wane to some extent over time. This particular theory of ``behaviour fatigue" was part of the rationale that drove a delay in implementing lockdown measures at the beginning of the COVID-19 pandemic in England, although the concept had limited supporting evidence \cite{harvey2020}. There have been some mathematical modelling studies done which have incorporated this type of behaviour dynamic \cite{juher2015, kassa2018, rypdal2020, meacci2021}. However, to our knowledge it has yet to be incorporated into any data-driven transmission models. By including a term representing fatigue (or to phrase more generally, a decrease in adherence) in a BC-ILM, we could investigate whether there is evidence for the existence of such an effect in an epidemic data set. Additionally, if behavioural fatigue or a similar phenomenon does exist, accounting for it in the model could improve accuracy, particularly if longer-term modelling or forecasting are of interest. 

\subsection*{Acknowledgements}
This work was supported by an University of Calgary Eyes High Doctoral Recruitment Scholarship, an Alberta Innovates Graduate Student Scholarship for Data-Enabled Innovation, a Natural Sciences and Engineering Research Council of Canada (NSERC) Postgraduate Scholarship-Doctoral, and the NSERC Discovery Grants Program.

\bibliographystyle{apalike} 


\newpage
\begin{appendix}

\section{Parameter estimates}

\begin{figure}[H]
\centering
\includegraphics[scale=0.65]{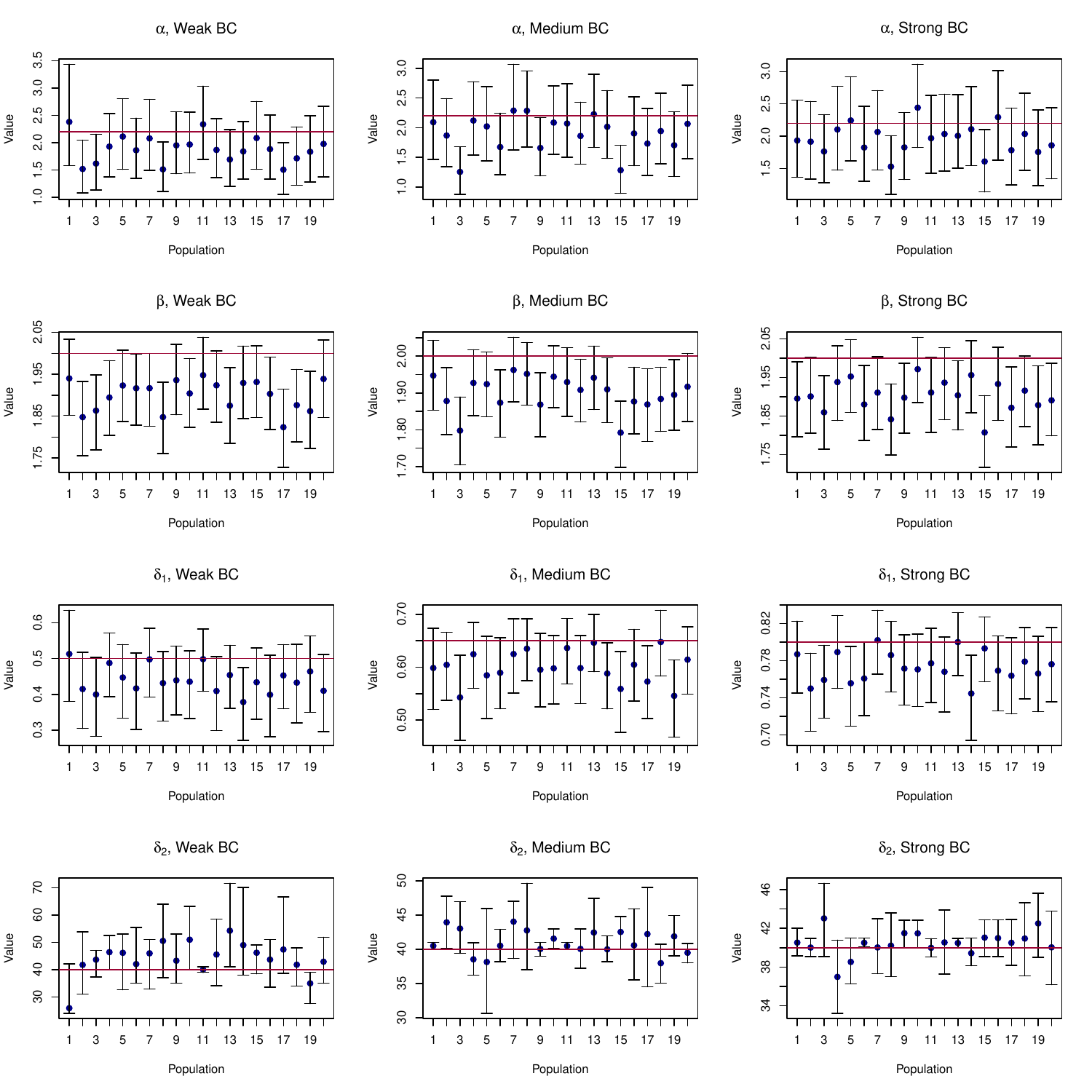}
\caption{95\% HPDIs and posterior medians (blue circles) for Model 1A parameters. Horizontal lines represent the true parameter values. } \label{fig:paramsmod1a}
\end{figure}

\begin{figure}[H]
\centering
\includegraphics[scale=0.65]{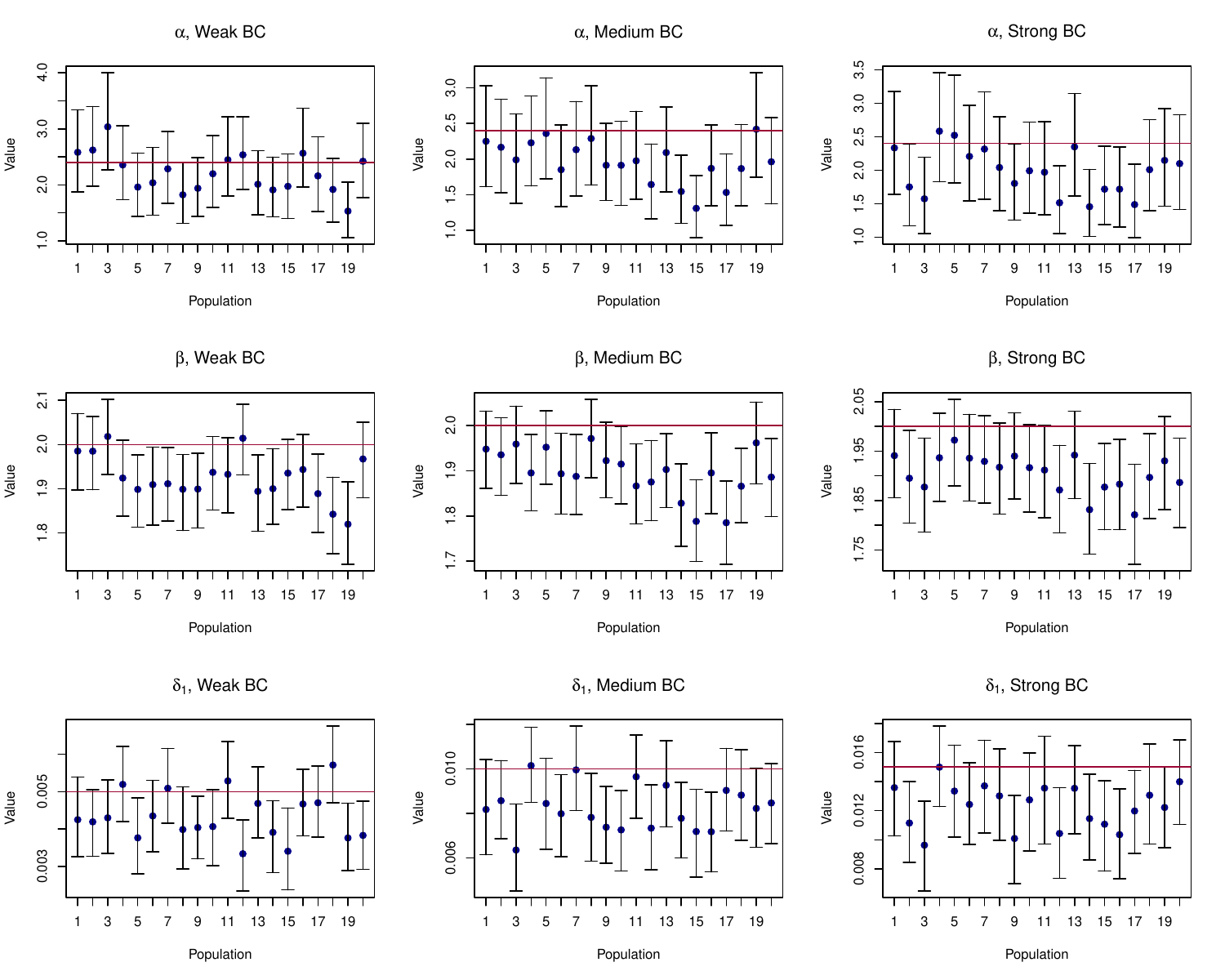}
\caption{95\% HPDIs and posterior medians (blue circles) for Model 2A parameters. Horizontal lines represent the true parameter values.} \label{fig:paramsmod2a}
\end{figure}

\begin{figure}[H]
\centering
\includegraphics[scale=0.65]{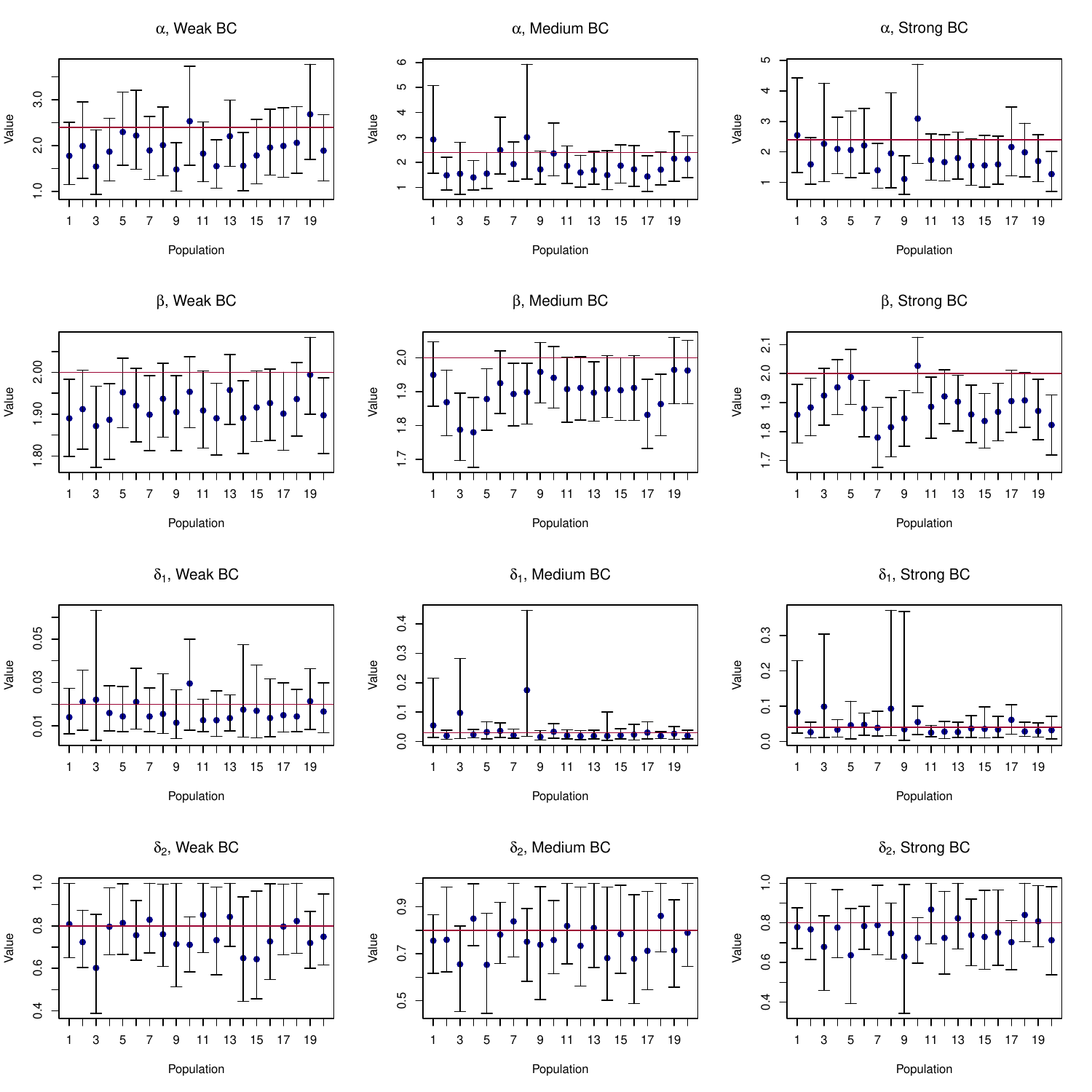}
\caption{95\% HPDIs and posterior medians (blue circles) for Model 3A parameters. Horizontal lines represent the true parameter values.} \label{fig:paramsmod3a}
\end{figure}

\begin{figure}[H]
\centering
\includegraphics[scale=0.65]{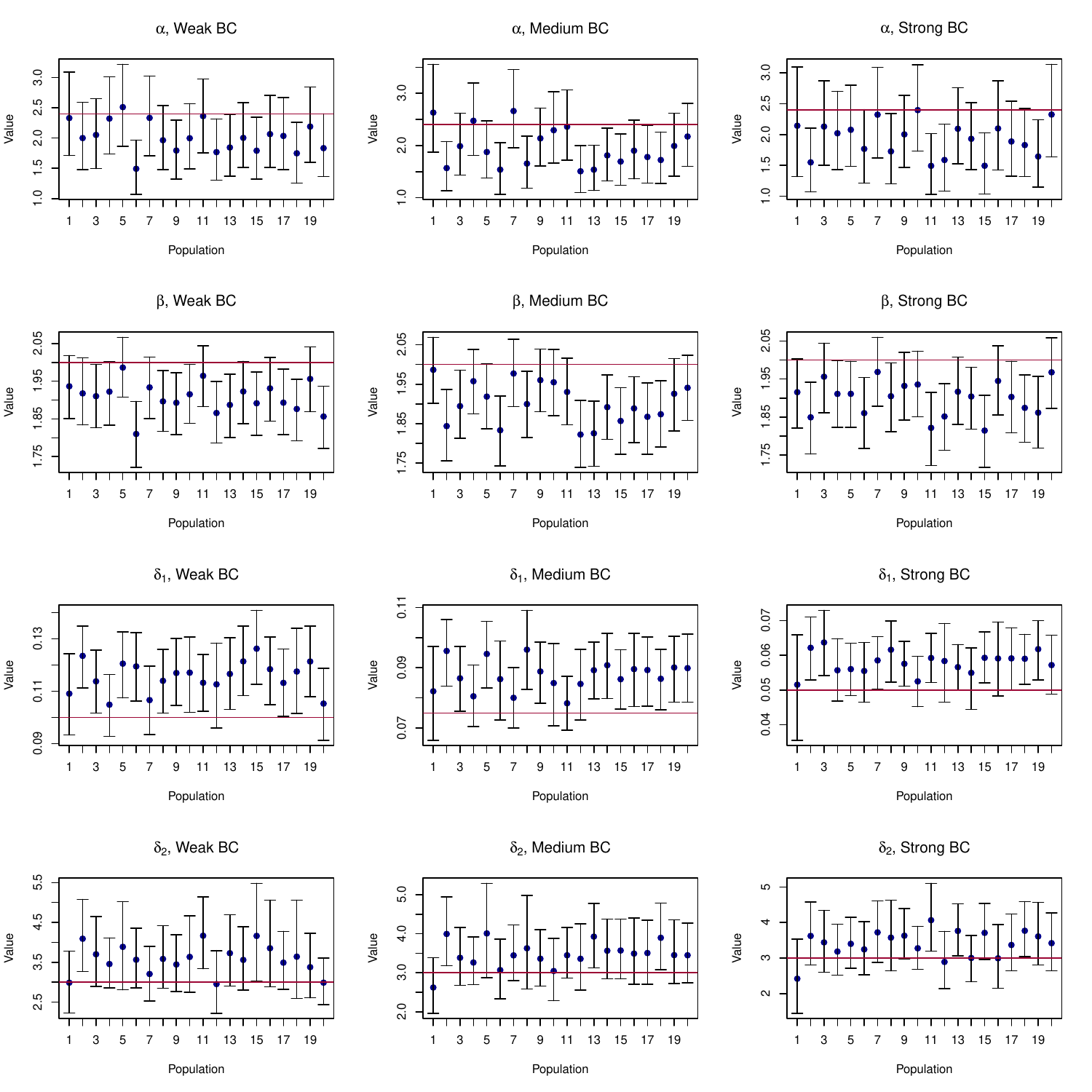}
\caption{95\% HPDIs and posterior medians (blue circles) for Model 4A parameters. Horizontal lines represent the true parameter values. } \label{fig:paramsmod4a}
\end{figure}

\section{Posterior predictive distributions}

\begin{figure}[H]
\centering
\includegraphics[scale=0.4]{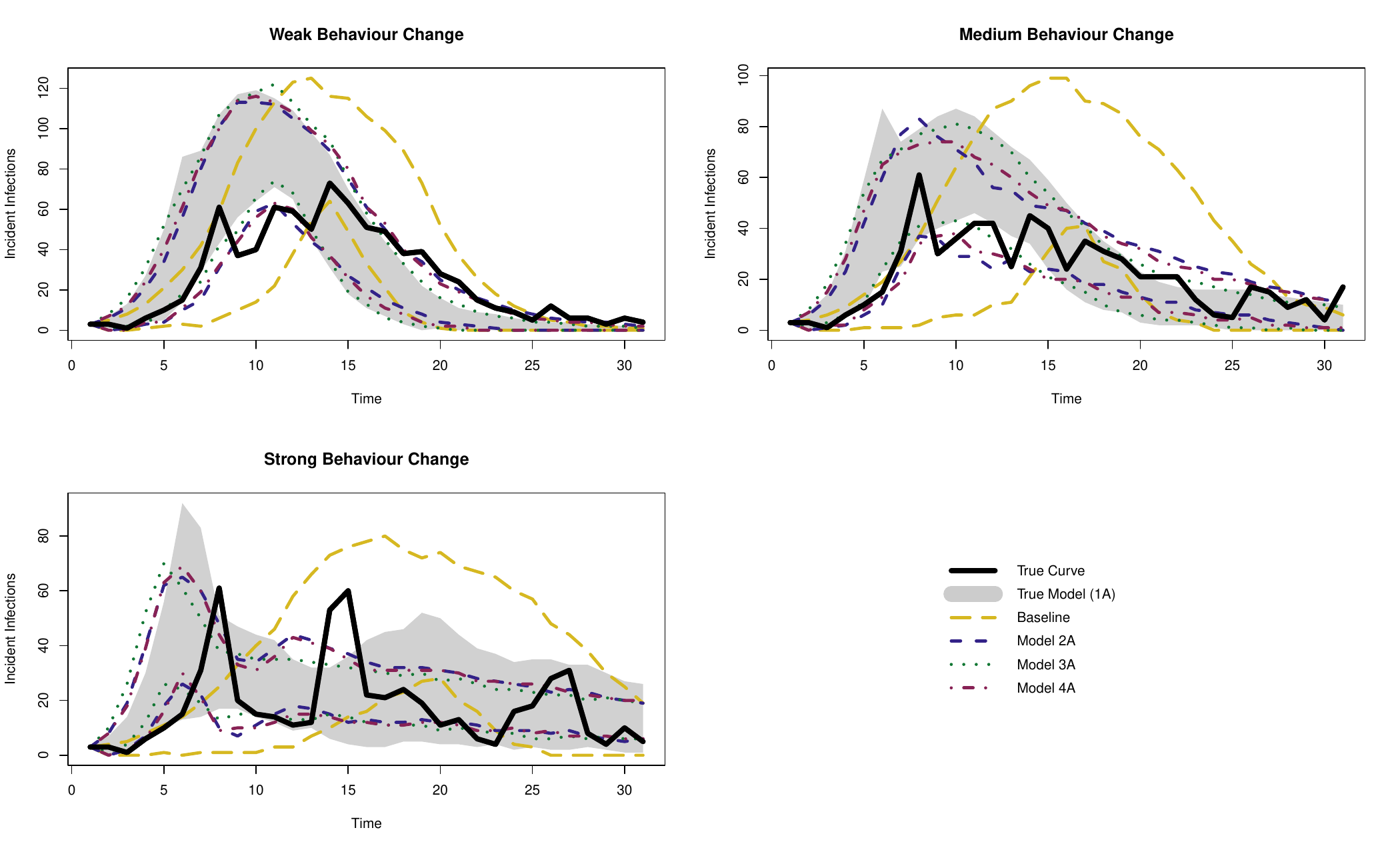}
\caption{95\% HPDIs for the epidemic curve under the true and misspecified models for a single representative epidemic simulated from Model 1A.  } \label{fig:curvehpdsmod1a}
\end{figure}

\begin{figure}[H]
\centering
\includegraphics[scale=0.4]{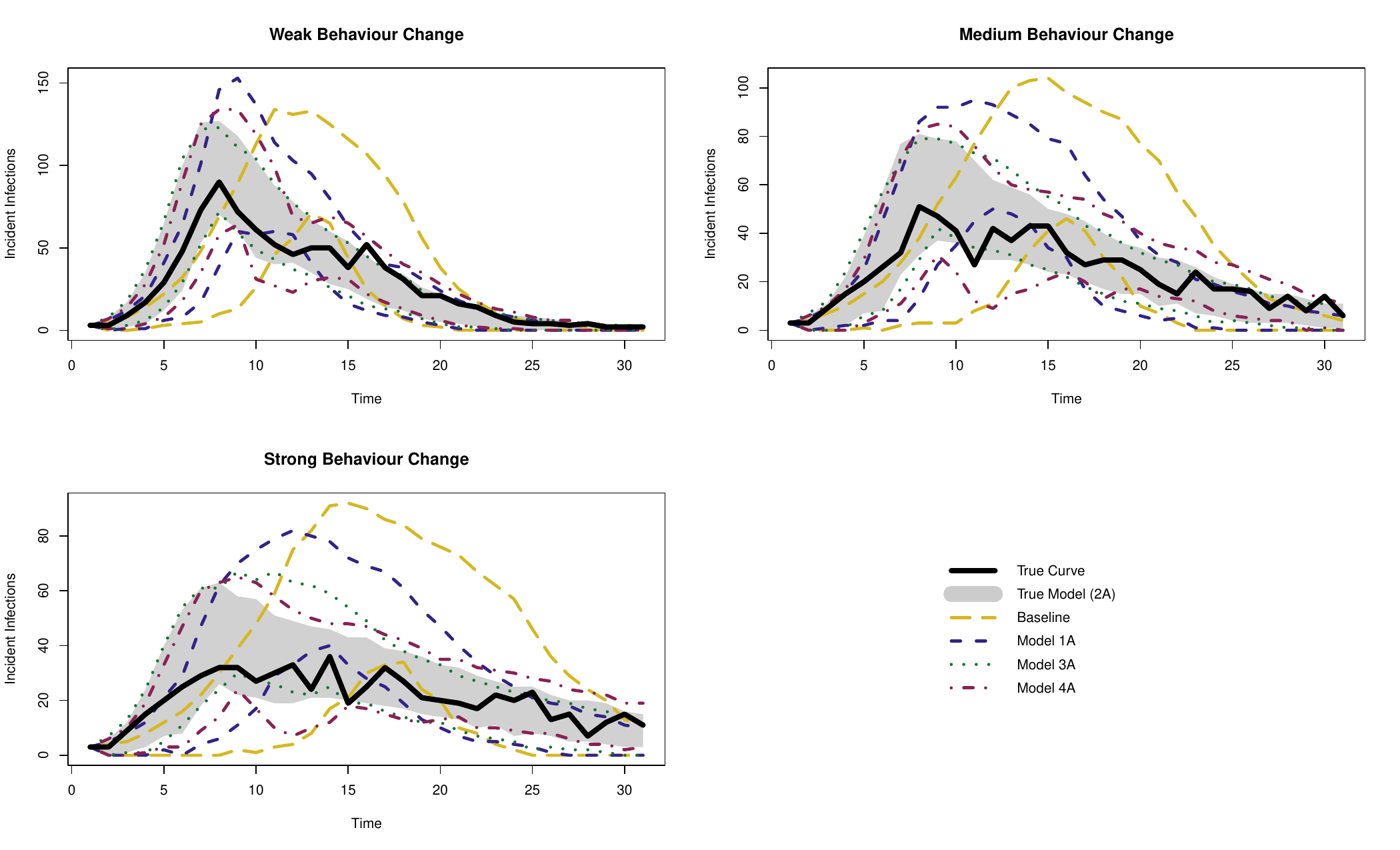}
\caption{95\% HPDIs for the epidemic curve under the true and misspecified models for a single representative epidemic simulated from Model 2A.  } \label{fig:curvehpdsmod2a}
\end{figure}

\begin{figure}[H]
\centering
\includegraphics[scale=0.4]{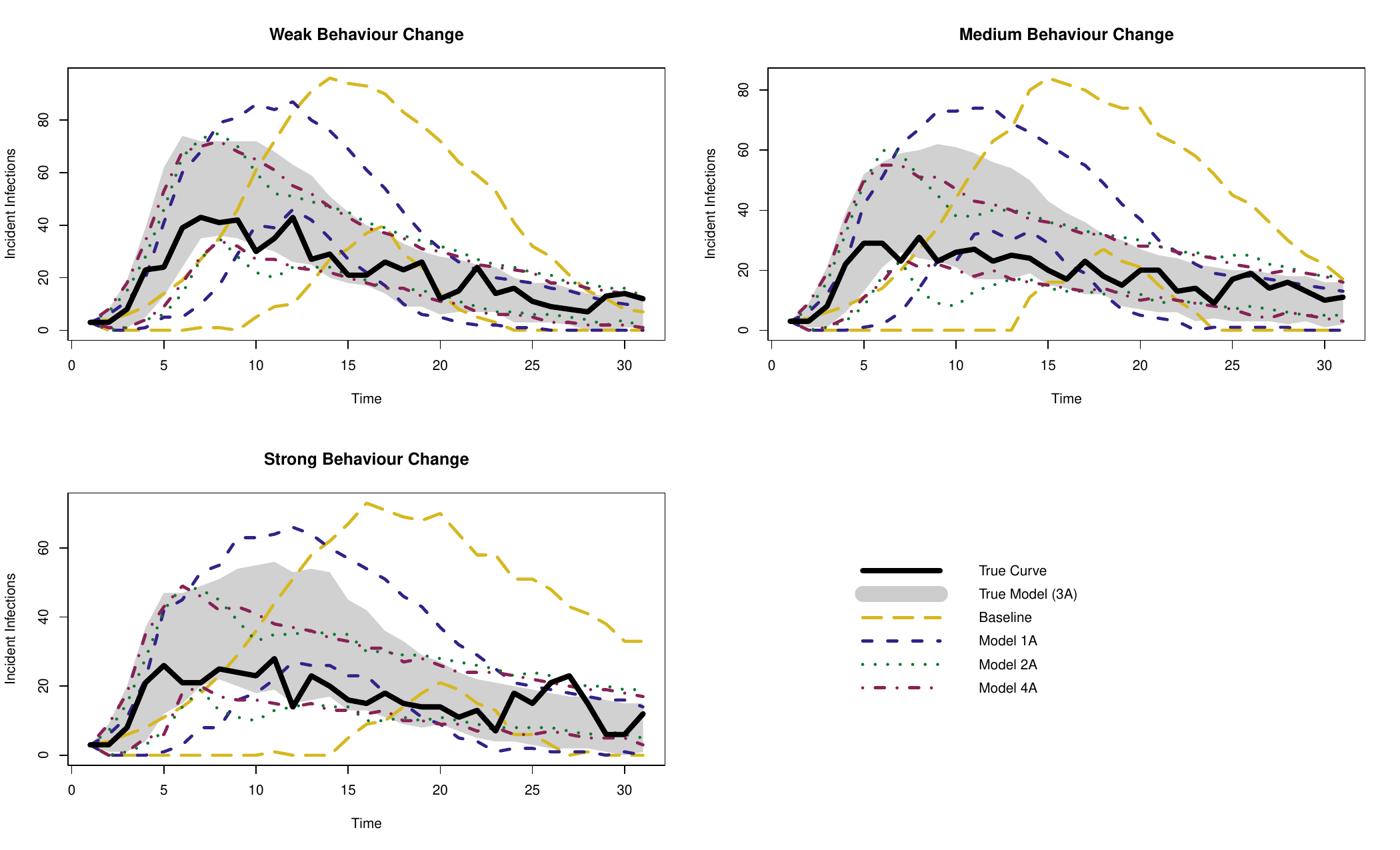}
\caption{95\% HPDIs for the epidemic curve under the true and misspecified models for a single representative epidemic simulated from Model 3A.  } \label{fig:curvehpdsmod3a}
\end{figure}

\begin{figure}[H]
\centering
\includegraphics[scale=0.4]{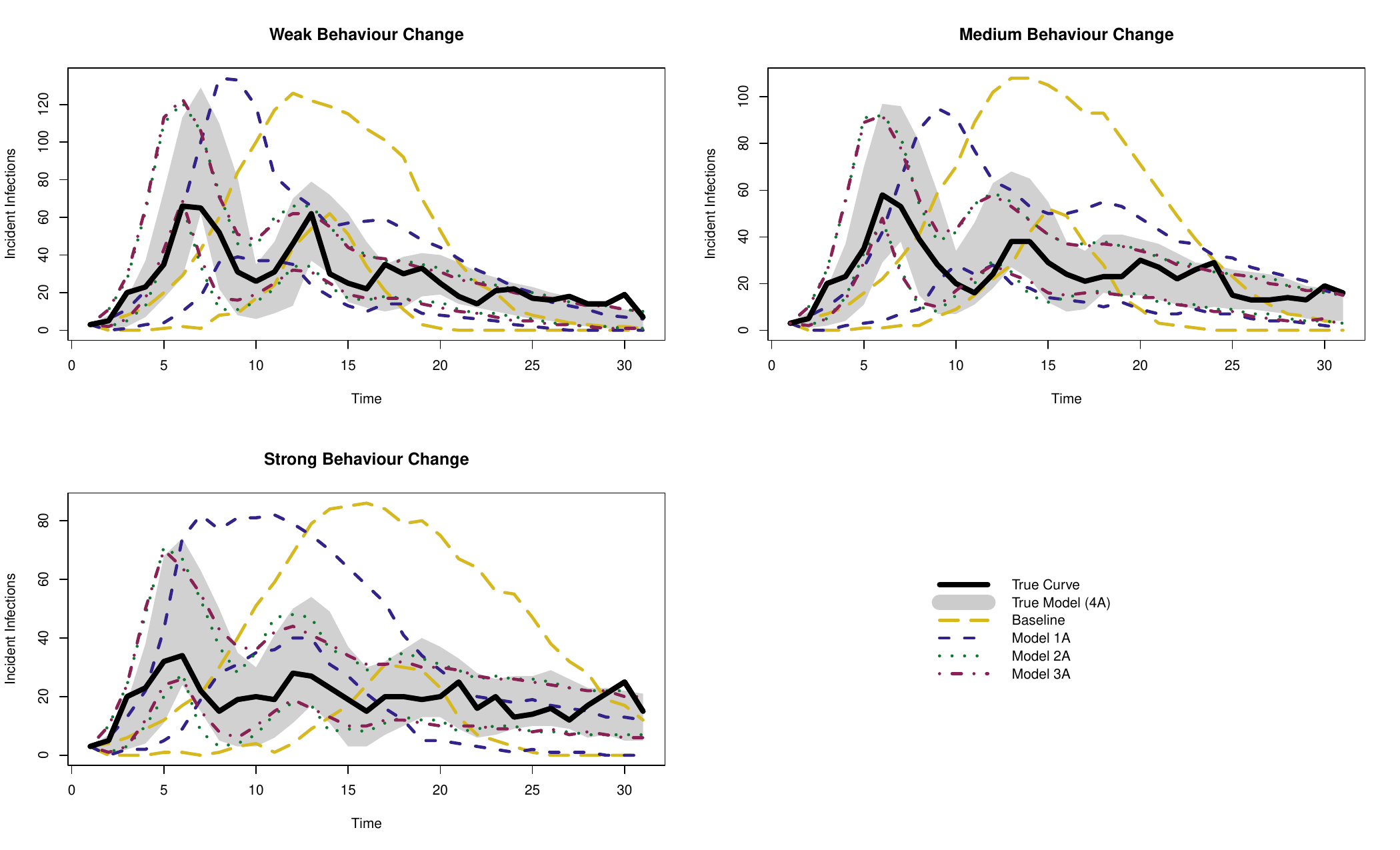}
\caption{95\% HPDIs for the epidemic curve under the true and misspecified models for a single representative epidemic simulated from Model 4A.  }
\label{fig:curvehpdsmod4a}
\end{figure}
\end{appendix}

\end{document}